\documentclass[twocolumn,aps]{revtex4}
\usepackage[english]{babel}

\usepackage{amsmath,amssymb,amsxtra,amsfonts}
\usepackage{graphicx}
\usepackage{multirow}
\usepackage{dcolumn}
\usepackage{txfonts}
\def\bea{\begin{eqnarray}}
\def\eea{\end{eqnarray}}
\def\be{\begin{equation}}
\def\ee{\end{equation}}

\begin{document}

\title{What do we know about cosmography}
\author{Ming-Jian Zhang$^{a}$}\email[Corresponding author : ]{zhangmj@ihep.ac.cn}
\author{Hong Li$^{a}$}
\author{Jun-Qing Xia$^{b}$}
\affiliation{$^a$Key Laboratory of Particle Astrophysics, Institute of High Energy Physics,
Chinese Academy of Science, P. O. Box 918-3, Beijing 100049, China}
\affiliation{$^b$Department of Astronomy, Beijing Normal University, Beijing 100875, China}

\begin{abstract}
In the present paper, we investigate the cosmographic problem using the bias-variance tradeoff. We find that both the $z$-redshift and $y=z/(1+z)$-redshift can present a small bias estimation. It means that the cosmography can describe the supernova data more accurately. Minimizing risk, it suggests that cosmography up to the 2 order is the best approximation. Forecasting the constraint from future measurements, we find that future supernova and redshift drift can significantly improve the constraint, thus having a potential to solve the cosmographic problem. We also exploit the values of cosmography on the deceleration parameter and equation of state of dark energy $w(z)$. We find that supernova cosmography cannot give stable estimations on them. However, many useful information were obtained, such as that the cosmography favors a complicated dark energy with varying $w(z)$, and the derivative $\textmd{d}w/\textmd{d}z <0$  for low redshift. The cosmography is helpful to model the dark energy.

\end{abstract}

\maketitle

\section{Introduction}
\label{introduction}

Cosmic accelerating expansion is a landmark cosmological discovery in recent decades. Till now, a number of dynamical mechanisms have been proposed to explain this mysterious cosmological phenomenon. However, its natural essence is still not known to us. The theoretical attempts include dark energy, or modified gravity, or violation of cosmological principle. Thereinto, the first paradigm believes that an exotic cosmic component called dark energy probably exists in a form of the cosmological constant \cite{carroll1992cosmological}, or scalar field
\cite{caldwell2003phantom,feng2005wang} and possesses a negative pressure to drive the cosmic acceleration. The modified gravities do not need an exotic component but a modification to the general relativity theory
\cite{barrow1988inflation,dvali20004d}. While violation of cosmological principle is usually in form of the inhomogeneous Lema\^{\i}tre-Tolman-Bondi
void model
\cite{lemaitre1933univers,tolman1934effect,bondi1947spherically}.

Different from above dynamical templates, cosmic kinematics is a more moderate approach in understanding this acceleration. It only highlights a homogeneous and isotropic universe at the large scale. In this family, kinematic parameters independent of the cosmic dynamical models become very essential. For example, the scale factor $a(t)$ directly describes how the universe evolves over time. The deceleration factor can immediately map the decelerating or accelerating expansion of the universe. Collecting some kinematic parameters, the authors in Refs. \cite{chiba1998luminosity,visser2004jerk} created the cosmography via Taylor expansion of the luminosity distance over the redshift $z$. Mathematically, this expansion should be performed near a small quantity, i.e. the low redshift. Using the standard complex variable theory, Catto\"{e}n and Visser \cite{cattoen2007hubble} demonstrated that convergence radius of Taylor expansion over redshift $z$ is at most $|z|=1$. For high redshift $z>1$, it fails to convergence. Nevertheless, many observations focus on the high redshift region. For example, the supernova in joint light-curve analysis (JLA) compilation can span the redshift region up to 1.3; the cosmic microwave background (CMB) even can retrospect to the very early universe at $z \sim 1100$. To legitimate the expansion at high redshift, they introduced an improved redshift parametrization $y=z/(1+z)$ \cite{cattoen2007hubble}. Thus, cosmography in the $y-$based expansion is mathematically safe and useful, because of $0<y<1$, even for the high redshift. Later, some other fashions of redshift were also proposed \cite{aviles2012cosmography}.

When confronting with observational data, the cosmography study encounter some difficulties. Initially, the SNIa data were used to fit with the cosmography \cite{cattoen2008cosmographic,capozziello2011comprehensive}. Then some auxiliary data sets  \cite{xia2012cosmography} were also considered. The output indicated that fewer series truncation lead to smaller errors but worse estimation; and more terms lead to more accurate approximation but bigger errors. That is, cosmography is in the dilemma between accuracy and precision. The crisis naturally turns  to the question of where is the ``sweet spot", i.e., the most optimized series truncation. In previous work \cite{cattoen2008cosmographic,vitagliano2010high,xia2012cosmography}, they found that estimation up to the snap term is meaningless in the light of \textit{F}-test. For the $y$-redshift, it presents bigger errors of the parameters \cite{cattoen2008cosmographic}. In spite of different observational data sets were used, most of the results were consistent. Recent work \cite{lazkoz2013bao} investigated the cosmography using the baryon acoustic oscillations (BAO) only. From the simulated Euclid-like BAO survey, they found that future BAO observation also favored a best cosmography with a jerk term. Because it only requires the homogeneity and isotropy of the universe, cosmography was frequently used to deduce or test the cosmological models. Recently, it was used to test the $\Lambda$CDM model \cite{busti2015cosmography}, but it turned out that the parameter $j_0 \neq 1$ is ambiguous for different orders of expansion, which is not enough to test it. Reconstructing dark energy in $f(R)$ gravities, they found that there exists extra free parameters, which cannot be constrained by the cosmography. The analysis was based on the mock data generated by a unified error of magnitude $\sigma_{\mu}=0.15$. In the following text, we will test the constraint of these mock data with flat errors. Following the work in Ref. \cite{busti2015cosmography}, these mock data were generated assuming the same redshift distribution as the Union 2.1 catalogue \cite{suzuki2012hubble}, but under the fiducial model from the best-fit ones by JLA data.

Although the cosmography has been widely investigated, there still left a lot of questions. On the one hand, we do not present a repetitive work using more data, but numerically excavate more detailed information about the accuracy and precision in convergence problem. The new approach we use is the bias-variance tradeoff. Moreover, we will try to investigate whether future measurement can solve the serious convergence issue. On the other hand, before the use of cosmography, we should make certain what information it can provide and what it cannot provide.

Although many types of observational data were used to fit the cosmography, our goal in this paper is to understand the convergence problem from another side, i.e., the geometric or dynamical measurement. Future surveys with high-precision may present a different constraint. To understand above questions, we need the help of future WFIRST-like supernova observation and a dynamical  survey redshift drift. Different from the geometric measurement, the redshift drift is desired to measure the secular variation of $\dot{a}(t)$ \cite{sandage1962change}. In contrast, geometric observation usually measure an integral of  $\dot{a}(t)$. Interestingly, this concept is also independent of any cosmological model, requiring only the Friedmann-Robertson-Walker universe. Taking advantage of the capacity of E-ELT \cite{liske2008cosmic,liske2008elt,corasaniti2007exploring}, numerous works agreed that this future probe could provide excellent contribution to understand the cosmic dynamics, such as the dark energy
\cite{zhang2014observational,geng2015redshift} or modified
gravity models \cite{li2013probing}. More importantly, it can be extended to test the fundamental Copernican principle \cite{uzan2008time} and the cosmic  acceleration \cite{yu2014method}. However, study of the redshift drift on kinematics has been scarce.

This paper is organized as follows: In Section~\ref{cosmography}, we introduce the
cosmography. And in Section~\ref{observation} we present the observational data. According to the goals introduced above, we analyze the problem of cosmography in Section \ref{problem}, and explore its values in Section \ref{values}. Finally, in Section \ref{conclusion} conclusion and discussion are drawn.

\section{Cosmography}
\label{cosmography}

Cosmography is an artful combination of kinematic parameters via the Taylor expansion with a hypothesis of large-scale homogeneity and isotropy. In this framework, introduction of the interested cosmographic parameters is appropriate.

Hubble parameter
\begin{equation}  \label{hubble parameter}
   H(t)= + \frac{1}{a} \frac{da}{dt}
\end{equation}
accurately connects the cosmological models with observational data.

Deceleration parameter
\begin{equation}  \label{deceleration}
   q(t)= - \frac{1}{a} \frac{d^2 a}{dt^2} \left[\frac{1}{a} \frac{da}{dt}\right]^{-2}
\end{equation}
directly represents the decelerating or accelerating expansion of the universe.

Jerk parameter
\begin{equation}  \label{jerk}
   j(t)= + \frac{1}{a} \frac{d^3 a}{dt^3} \left[\frac{1}{a} \frac{da}{dt}\right]^{-3}
\end{equation}
and snap parameter
\begin{equation}  \label{snap}
   s(t)= + \frac{1}{a} \frac{d^4 a}{dt^4} \left[\frac{1}{a} \frac{da}{dt}\right]^{-4}
\end{equation}
are often used as a geometrical diagnostic of dark energy models \cite{sahni2003statefinder,alam2003exploring}. An important feature should be announced is that jerk has being a traditional tool to test the spatially flat cosmological constant dark energy model in which $j(z)=1$ all time.

Lerk parameter
\begin{equation}  \label{lerk}
   l(t)= + \frac{1}{a} \frac{d^5 a}{dt^5} \left[\frac{1}{a} \frac{da}{dt}\right]^{-5}
\end{equation}
is an higher order parameter to indicate the cosmic expansion.

With above preparation, Hubble parameter in the cosmography can be expressed as \cite{chiba1998luminosity,capozziello2011comprehensive}
\bea  \label{Hz taylor}
H(z) &=& H_0 + \frac{dH}{dz}\Big|_{z=0} z + \frac{1}{2!}\frac{d^2H}{dz^2}\Big|_{z=0} z^2 + \frac{1}{3!}\frac{d^3H}{dz^3}\Big|_{z=0} z^3 +  \cdots                  \nonumber\\
     &=& H_0\Big[1 + (1+q_0) z +\frac{1}{2}(-q_0^2+j_0)z^2              \nonumber\\
     &+& \frac{1}{6}(3q_0^2+3q_0^3-4 q_0 j_0-3 j_0 -s_0)z^3             \nonumber\\
     &+& \frac{1}{24}(-12q_0^2-24q_0^3-15q_0^4+32q_0 j_0 +25 q_0^2 j_0  \nonumber\\
     &+& 7 q_0 s_0  + 12 j_0-4 j_0^2+ 8 s_0 + l_0) z^4 \Big]  +  \cdots
\eea
where the subscript ``0" indicates cosmographic parameters evaluated at the present epoch. According to the differential relations with Hubble parameter, luminosity distance in the cosmography study can be conveniently expressed as  \cite{cattoen2007hubble,capozziello2011comprehensive}
\begin{equation}  \label{modulus z}
   d_L^{\textmd{cos}}(z) = z + \mathcal{C}_1 z^2 + \mathcal{C}_2 z^3 + \mathcal{C}_3 z^4 + \mathcal{C}_4 z^5 ,
\end{equation}
where
\begin{eqnarray}
   \mathcal{C}_1 &=& \frac{1}{2}(1-q_0)                                          \nonumber\\
   \mathcal{C}_2 &=& -\frac{1}{6}(1- q_0-3q_0^2+ j_0)                        \nonumber\\
   \mathcal{C}_3 &=& \frac{1}{24}(2-2 q_0-15q_0^2-15 q_0^3+5 j_0+10 q_0 j_0 +s_0) \nonumber\\
   \mathcal{C}_4 &=& \frac{1}{120}(-6+6q_0+81q_0^2+165q_0^3+105q_0^4   \nonumber\\
   &+&10j_0^2-27j_0-110q_0 j_0-105q_0^2 j_0-15q_0 s_0                      \nonumber\\
   &-&11s_0-l_0) ,
\end{eqnarray}
As introduced above, cosmography at high redshift $z>1$ fails to converge. To solve this trouble, a $y$-redshift hence introduced \cite{cattoen2007hubble}
\be
y = \frac{z}{1+z}.
\ee
For the new redshift, we can simplify it as $y=1-a(t)$. Obviously, it is $0<y<1$ for the current observational data. One benefit from the $y$-redshift is that it can extend the expansion to high redshift region. This is important for the cosmography study. The reason is that cosmography in Eq. \eqref{modulus z} is theoretically valid for redshift $z<1$. While with the import of $y$-redshift, many observational data such as supernova with higher redshift and even CMB, can be used to fit and study the cosmography. For example, it is reduced to $y=0.56$ for the supernova at max redshift $z=1.3$ in JLA compilation. Moreover, its value is $y=0.999$ which guarantees the safe use of the early CMB data. As described in Ref. \cite{cattoen2007hubble}, it even can extrapolate back to the big bang. The other physical significance is $y$ redshift also can back to the future universe, but breaks down at $y=-1$. In the $y$-redshift space,  the luminosity distance is
\begin{equation}  \label{modulus y}
   d_L^{\textmd{cos}}(y) = y + \mathcal{C}_1 y^2 + \mathcal{C}_2 y^3 + \mathcal{C}_3 y^4 + \mathcal{C}_4 y^5 ,
\end{equation}
with
\begin{eqnarray}
   \mathcal{C}_1 &=& \frac{1}{2}(3-q_0)                                         \nonumber\\
   \mathcal{C}_2 &=& \frac{1}{6}(11-5 q_0+3q_0^2- j_0)                         \nonumber\\
   \mathcal{C}_3 &=& \frac{1}{24}(50- 26q_0+21 q_0^2-15 q_0^3-7 j_0+10 q_0 j_0 +s_0)  \nonumber\\
   \mathcal{C}_4 &=& \frac{1}{120}(274-154q_0+141q_0^2-135q_0^3+105q_0^4        \nonumber\\
   &+&10j_0^2-47j_0+90q_0 j_0-105q_0^2 j_0-15q_0 s_0                         \nonumber\\
   &+&9s_0-l_0) .
\end{eqnarray}
In following analysis, one we should do is to test the improvement of $y$-redshift.

In our cosmographic study, we need the help of dynamical redshift drift. The story should start from the redshift.

In an expanding universe, we observe at
time $t_0$ a signal emitted by a source at $t_{\mathrm{em}}$.  The
source's redshift can be represented through the cosmic scale factor
    \be
    z(t_0) = \frac{a(t_0)}{a(t_{\mathrm{em}})} -1.
    \ee
Over the observer's time interval $\Delta t_0$, the source's
redshift  becomes
    \be
    z(t_0+\Delta t_0) = \frac{a(t_0+\Delta t_0)}{a(t_{\mathrm{em}}+\Delta t_{\mathrm{em}})}-1,
    \ee
where $\Delta t_{\mathrm{em}}$ is the time interval-scale for the
source to emit another signal. It should satisfy  $\Delta
t_{\mathrm{em}} = \Delta t_0 / (1+z)$. As a consequence, the observed
redshift variation of the source is
    \be   \label{Dz}
    \Delta z=\frac{a(t_0+\Delta t_0)}{a(t_{\mathrm{em}}+\Delta t_{\mathrm{em}})}
    - \frac{a(t_0)}{a(t_{\mathrm{em}})}.
    \ee
Taking the first order approximation to Eq. \eqref{Dz}, physical interpretation of redshift drift can be exposed in
    \be  \label{Dz approximation}
    \Delta z \approx \left[ \frac{\dot{a}(t_0) - \dot{a}(t_{\mathrm{em}})}{a(t_{\mathrm{em}})} \right] \Delta t_0 ,
    \ee
where dot denotes the derivative with respect to cosmic time. Obviously, we should note that the secular redshift drift monitor a variation of $\dot{a}$ during the evolution of the universe. For the distance measurement, it commonly extracts information content via the integral of a variant of $\dot{a}$. Theoretically, the Hubble parameter, a function of $\dot{a}$ may be more effective to probe the cosmic expansion information. However, its acquisition in observational cosmology currently is \textit{indirect}  from the differential ages of galaxies
\cite{jimenez2008constraining,simon2005constraints,stern2010cosmic},
from the BAO peaks in the galaxy power
spectrum \cite{gaztanaga2009clustering,moresco2012improved}, or from
the BAO peaks using the Ly$\alpha$ forest of quasar (QSO)
\cite{delubac2013baryon}. For the redshift drift, we note that it is a direct measurement to the cosmic expansion and can be come true via multiple methods \cite{loeb1998direct}.

In terms of the Hubble parameter $H(z)=\dot{a}(t_{\mathrm{em}})/a(t_{\mathrm{em}})$, we simplify Eq. \eqref{Dz approximation} as
   \be  \label{drift define}
   \frac{\Delta z}{\Delta t_0}=(1+z)H_0 - H(z).
   \ee
What we should highlight is its independence of any prior and dark energy model. For this unique advantage, many analysis have demonstrated that the redshift drift is not only able to provide much stronger constraints on the dynamical cosmological models \cite{martinelli2012probing,li2013probing}, but also to solve some crucial cosmological problems \cite{moraes2011complementarity,zhang2014cosmic}, even allows us to test the Copernican principle \cite{uzan2008time}. Observationally, it is convenient to probe the spectroscopic velocity drift
\begin{equation}  \label{velocity define}
   \frac{\Delta \upsilon}{\Delta t_0} = \frac{c}{1+z} \frac{\Delta z}{\Delta t_0} ,
\end{equation}
which is of an order of several cm s$^{-1}$yr$^{-1}$. The signal is naturally accumulated with an increase of observational time $\Delta t_0$.

Taylor expansion tells us that the redshift drift should be
\begin{equation}  \label{velocity taylor}
   \frac{\Delta \upsilon}{\Delta t_0}=\frac{\Delta \upsilon}{\Delta t_0}\Big|_{z=0} + \frac{d}{dz}\left(\frac{\Delta \upsilon}{\Delta t_0}\right) \Big|_{z=0} z + \frac{1}{2!} \frac{d^2 }{dz^2}\left(\frac{\Delta \upsilon}{\Delta t_0}\right)\Big|_{z=0} z^2 + \cdots
\end{equation}
Using the Taylor series of Hubble parameter in Eq. \eqref{Hz taylor}, we can put the Eq. \eqref{velocity taylor} into practice
\begin{eqnarray}  \label{velocity z}
   \frac{\Delta \upsilon}{\Delta t_0} (z) &=& c H_0 \Big[  -q_0 z + \frac{1}{2} (2q_0 +q_0^2 -j_0)  z^2  \nonumber\\
   &+&  \frac{1}{6} (-6q_0 -6q_0^2 -3q_0^3 +4q_0j_0 + 6j_0  +s_0)z^3 \nonumber\\
   &+&  \frac{1}{24}(24q_0+36q_0^2+36q_0^3+15q_0^4-48q_0 j_0-25q_0^2 j_0  \nonumber\\
   &-& 7q_0 s_0-36j_0+4j_0^2-12s_0-l_0)z^4 \Big] .
\end{eqnarray}
For the $y$-redshift, it is simplified as
\begin{eqnarray}   \label{velocity y}
   \frac{\Delta \upsilon}{\Delta t_0} (y)&=& cH_0 \Big[ -q_0 y + \frac{1}{2} (q_0^2-j_0)  y^2 \nonumber\\
   &+& \frac{1}{6}(-3q_0^3+4q_0 j_0 +s_0) y^3 \nonumber\\
   &+& \frac{1}{24}(15q_0^4-25q_0^2 j_0+ 4j_0^2-7q_0 s_0-l_0) y^4  \Big] .
\end{eqnarray}
Recently, Taylor expansion of the redshift drift was also provided in the varying speed of light cosmology \cite{balcerzak2014redshift}. One can reduce it from the non-mainstream scenario to classical case. In this paper, we mainly use it to provide a numerical constraint on the cosmography, to test its constraint power on the cosmic kinematics.

\section{Observational data}
\label{observation}

In this section, we introduce the related data in our calculation. Current data we use are the canonical distance modulus from JLA compilation. In order to test whether future SNIa observation can alleviate or terminate the tiresome convergence problem,  we produce some mock data by the Wide-Field InfraRed Survey Telescope-Astrophysics Focused Telescope Assets (WFIRST-AFTA) \footnote{http://wfirst.gsfc.nasa.gov/}. The dynamical redshift drift is forecasted by the E-ELT. The parameters can be estimated through a Markov chain Monte Carlo method, by modifying the publicly available code \verb"CosmoMC" \cite{lewis2002cosmological}. As introduced in Section \ref{cosmography}, the cosmography is independent of dynamical models. Therefore, we fix the background variables, and relax the cosmographic parameters as free parameters in our calculation.

\subsection{Current supernova}  \label{current SN}

One important reason of why the supernova data were widely used is its extremely plentiful resource. In this
paper, we use the latest supernova JLA compilation of 740 dataset from the SDSS and SNLS \cite{betoule2014improved}. The data are usually presented as tabulated distance modulus with errors. In this catalog, the redshift spans $z<1.3$, and about 98.9\% samples are in the redshift region $z<1$. In our calculation, we also consider all the covariance matrix.

\subsection{Future supernova}  \label{future SN}

In cosmology study, forecasting the constraint of future observations on the cosmological model is quite useful for theory research. Estimation to the uncertainty of the observational variable is a core matter. In previous cosmography study, one usually makes several mock data from a conceptual telescope or satellite \cite{xia2012cosmography}, or extrapolation from current observational data \cite{2015ApJ...814....7B}. To be more reliable, in the present paper, we plan to use a living program. The WFIRST-2.4 not only stores tremendous potential on some key scientific program, but also enables a survey with more supernova in a more uniform
redshift distribution. One of its science drivers is to measure the cosmic expansion history. According to the updated report by Science Definition Team \cite{spergel2015wide}, we obtain 2725 SNIa over the region $0.1<z<1.7$ with a bin $\Delta z=0.1$ of the redshift.

The photometric measurement error per supernova is $\sigma_{\textmd{meas}} = 0.08$ magnitudes. The
intrinsic dispersion in luminosity is assumed as $\sigma_{\textmd{int}} = 0.09$ magnitudes (after correction/matching for light curve shape and spectral properties). The other contribution to statistical errors is gravitational lensing magnification, $\sigma_{\textmd{lens}} = 0.07 \times z$
mags. The overall statistical error in each redshift bin is then
\be
\sigma_{\textmd{stat}} = \left[(\sigma_{\textmd{meas}})^2 + (\sigma_{\textmd{int}})^2 + (\sigma_{\textmd{lens}})^2 \right]^{1/2} / \sqrt{N_i} ,
\ee
where $N_i$ is the number of supernova in the $i$-th redshift bin. According to the
estimation, a systematic error per bin is
\be
\sigma_{\textmd{sys}} = 0.01 (1+z) /1.8  .
\ee
Therefore, the total error per redshift bin is
\be
\sigma_{\textmd{tot}} = \left[(\sigma_{\textmd{stat}})^2 + (\sigma_{\textmd{sys}})^2 \right]^{1/2} .
\ee
In our simulation, the fiducial models are taken from the best-fit values by current supernova on the cosmographic models. We should note that although we have considered the various error sources, it is still difficult to provide the total covariance matrix of future WFIRST like current supernova data. It may inevitably underestimate the errors of cosmographic parameters. However, this forecast is helpful for us to study whether future observation can improve the convergence problem.

\subsection{Redshift drift} \label{future drift}

As suggested by Loeb \cite{loeb1998direct}, the redshift drift probe can come true via the wavelength shift of the QSO Ly$\alpha$ absorption lines, emission spectra of galaxies, and some other radio techniques. Thereinto, the ground-based largest optical/near-infrared
telescope E-ELT will prefer to provide continuous monitor from the Ly$\alpha$ forest in the spectra of
high-redshift QSOs \cite{pasquini2006codex}. These spectra are not
only immune from the noise of the peculiar motions relative to the
Hubble flow, but also have a large number of lines in a single
spectrum \cite{pasquini2005codex}.
According to the capability of E-ELT, the
uncertainties of velocity drift can be modelled as
\cite{pasquini2005codex,liske2008cosmic}
    \begin{equation}\label{velocity error}
\sigma_{\Delta \upsilon} = 1.35 \left( \frac{\textrm{S/N}}{2370}
\right)^{-1}\left(  \frac{N_{\scriptsize \textrm{QSO}}}{30}
\right)^{-1/2}\left( \frac{1+z_{\scriptsize \textrm{QSO}}}{5}
\right)^{q} \mathrm{cm/s},
\end{equation}
with $q=-1.7$ for $2<z<4$, or $q=-0.9$ for $z>4$, where the
signal-to-noise ratio S/N is assumed as 3000, the number of QSOs $N_{\scriptsize \textrm{QSO}}=30$ and
$z_{\scriptsize \textrm{QSO}}$ is the
redshift at $2<z<5$. Following previous works \cite{corasaniti2007exploring,martinelli2012probing,li2013probing,zhang2014observational}, we can obtain the mock data assumed to be uniformly distributed among the redshift bins: $z_{\scriptsize \textrm{QSO}}= [2.0, \: 2.8, \: 3.5, \: 4.2, \: 5.0]$
under the fiducial model from the best-fit ones by JLA data. With no specific declaration, the observational time is set as ten years.

\section{Problem of the cosmography}
\label{problem}

Convergence problem has always been a top priority in the cosmography study. According to the requirement of Taylor expansion, we will respectively perform related calculation for data at $z<1$ and $y<1$. In this section, we will analyze the convergence issue in current observational data, and forecast the constraint of future measurement. To ensure the physical meaning of constraint, we should apply a prior on the Hubble parameter
\[
H(z) >0
\]
in our calculation for both the $z$-redshift and $y$-redshift.

\subsection{Convergence issue in current data}

\begin{figure}
    \begin{center}
\includegraphics[width=8.0cm,height=7.0cm]{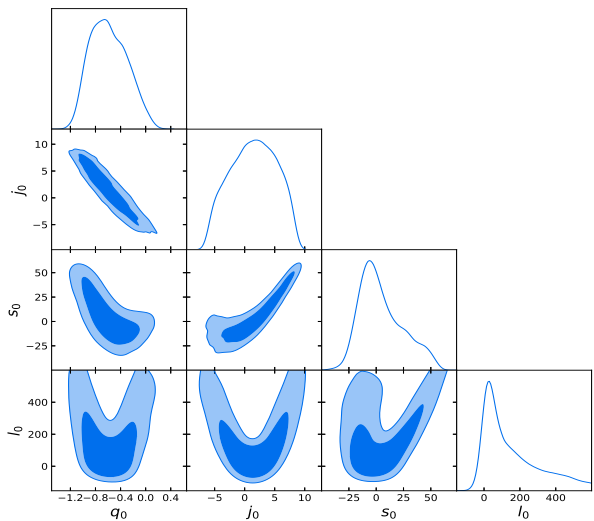}
    \end{center}
    \caption{\label{SNzcontour} Constraints on cosmographic parameters in the 4 order from the JLA compilation for redshift $z<1$.}
\end{figure}

\begin{figure}
    \begin{center}
\includegraphics[width=8.0cm,height=7.0cm]{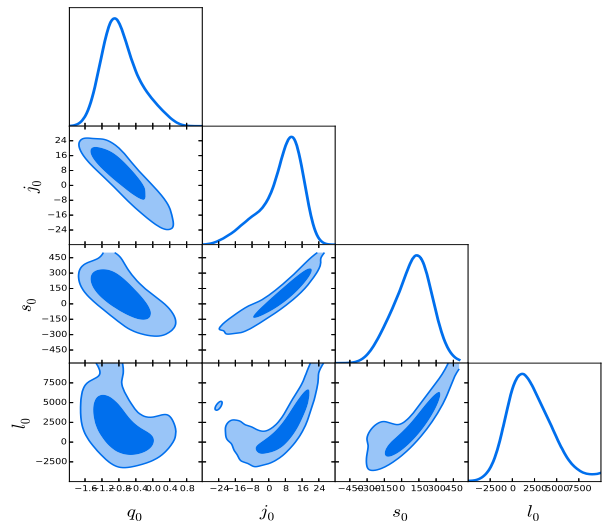}
    \end{center}
    \caption{\label{SNycontour} Constraints on cosmographic parameters in the 4 order from the JLA compilation for $y$-redshift.}
\end{figure}

\begin{figure}
\includegraphics[width=0.4\textwidth]{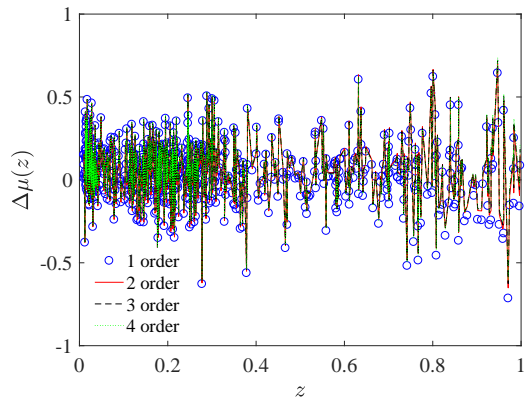}
\caption{Residuals between the cosmographic distance modulus with different orders and the observational SNIa data. The vertical coordinates $\Delta \mu (z) = \mu^{\textmd{cos}} (z) - \mu^{\textmd{obs}} (z)$ denotes the residuals. }
\label{residual}
\end{figure}

Using the JLA compilation, we obtain the cosmographic parameters up to the fourth order. We show the corresponding results in Figs. \ref{SNzcontour}, \ref{SNycontour}, \ref{residual} and Table \ref{JLA results}.

For the $z$-redshift, we can roughly distinguish the data and cosmographic models by residuals in Fig. \ref{residual}. On the one hand, most of the data locate at the low redshift, and fit well with the models. On the other hand, some of the data at high redshift present a little bigger residual with the cosmographic models. Thus, more low-$z$ data make the cosmography study more precise. From the constraints in Table \ref{JLA results}, we note that all of the constraints on parameter $q_0$ in $1\sigma$ confidence level are negative, which shows a recent accelerating expansion. However, some recent work try to find out a slowing down of the acceleration. Moreover, this novel phenomenon has attracted much attention \cite{shafieloo2009cosmic,cardenas2013cosmic,li2011examining,2013PhRvD..87d3502L,magana2014cosmic,2016ApJ...821...60W}, including the recent work \cite{2017MNRAS.469...47M}. Using the dark energy parameterizations, they found that the cosmic acceleration may has already peaked, and the expansion may be slowing down from the deceleration parameter $q_0>0$. In recent work \cite{seikel2012reconstruction,zhang2016test}, a model-independent analysis on this interesting subject was presented, using the powerful Gaussian processes technique. It was found that no slowing down is detected within $2\sigma$ C.L. from current data. Moreover, we analyzed the inconsistency in Ref.\cite{zhang2016test}. We further deduced what physical condition should be satisfied by the observational data \cite{zhang2017physical}. These results are consistent with the cosmographic constraint from JLA data.

Comparison between Fig. \ref{SNzcontour} and \ref{SNycontour} shows that degeneracies among the parameters in the $y$-redshift are similar as the $z$-redshift. The difference is that it provides much bigger errors on some parameters. Taking the parameter $l_0$ as an example, we find that its absolute value behaves an increasing trend. Moreover, its relative error in the $y$-redshift 2478.44/2056.97 is also bigger than that of the $z$-redshift 158.08/149.54. This is consistent with the result in previous work, namely, $y$-redshift brings worse constraints.

\begin{table*}[t]
\caption{ \label{JLA results} Constrained cosmographic parameters by the JLA compilation and mock data within $1\sigma$ confidence level.}
\small
$
\begin{array}{c c r@{ }l c r@{ }l c r@{ }l c r@{ }l}
\hline \hline
\multirow{2}{*}{data}  & \multicolumn{3}{c}{q_0}       & \multicolumn{3}{c}{j_0}
& \multicolumn{3}{c}{s_0} & \multicolumn{3}{c}{l_0} \\
& \multicolumn{1}{c}{\textmd{best fit}}   & \multicolumn{2}{c}{\textmd{mean}}
& \multicolumn{1}{c}{\textmd{best fit}}   & \multicolumn{2}{c}{\textmd{mean}}
& \multicolumn{1}{c}{\textmd{best fit}}   & \multicolumn{2}{c}{\textmd{mean}}
& \multicolumn{1}{c}{\textmd{best fit}}   & \multicolumn{2}{c}{\textmd{mean}}   \\ \hline
\multicolumn{9}{l}{\texttt{model}: z-\texttt{redshift}}   \\
\textmd{JLA}   &-0.46&-0.46&\pm0.04    &&&                  &&&                  &&&                   \\
                    &-0.45&-0.44&\pm0.09  &0.41&0.42&\pm0.24 &&&                 &&&                    \\
                    &-0.62&-0.50&\pm0.09  &1.93&1.06&\pm0.56   &1.77&0.50&\pm1.15 &&&        \\
                    &-0.67&-0.58&\pm0.29  &2.54&1.45&\pm3.82   &3.98&4.97&\pm20.35 & 19.90&149.54&\pm158.08    \\
  \hline
\multicolumn{9}{l}{\texttt{model}: y-\texttt{redshift}}  \\
\textmd{JLA}   &-0.72&-0.71&\pm0.07    &&&                  &&&                  &&&                   \\
                    &-0.42&-0.45&\pm0.20  &-0.15&0.14&\pm1.82 &&&                 &&&                    \\
                    &-0.86&-0.88&\pm0.44  &7.71&8.64&\pm7.92   &81.72&114.83&\pm113.52 &&&        \\
                    &-0.78&-0.74&\pm0.45  &6.36&6.16&\pm10.06   &69.69&89.06&\pm160.80 & 851.04&2056.97&\pm2478.44    \\
\textmd{WFIRST}     & &\sigma_{q_0}&=0.15  & &\sigma_{j_0}&=4.11   & &\sigma_{s_0}&=69.52 & &\sigma_{l_0}&=1146.30    \\
\textmd{redshift drift}     & &\sigma_{q_0}&=0.51  & &\sigma_{j_0}&=4.75   & &\sigma_{s_0}&=242.11 & &\sigma_{l_0}&=1299.17    \\
 \hline \hline
\end{array}
$


\end{table*}

In recent years, many work focus on the question of which series truncation fits the data best. In previous work \cite{cattoen2008cosmographic,vitagliano2010high,xia2012cosmography}, one introduced the \textit{F}-test to find this answer, by favoring one model, and assessing the other alternative model. Although it showed that expansion up to the jerk term is a better description for observational luminosity distance, the cosmographic problem is still vague. It is difficult for us to escape from the maze of cosmography in the accuracy and precision. We should underline that small error does not mean a credible description, and big error is not necessarily a bad thing. For further analysis upon this issue, we recommend the bias-variance tradeoff \cite{wasserman2006all}
\bea   \label{risk equation}
\textmd{risk} &=& \textmd{bias}^2 + \textmd{variance} \nonumber\\
              &=& \sum_{i=1}^N [\mu^{\textmd{cos}}(z_i) - \tilde{\mu}(z_i)]^2 + \sum_{i=1}^N \sigma^2 (\mu^{\textmd{cos}}(z_i))
\eea
where $\mu^{\textmd{cos}}(z_i)$ is the reconstructed cosmographic distance modulus in different series truncations, $\tilde{\mu}(z_i)$ is the fiducial value, $\sigma (\mu^{\textmd{cos}}(z_i))$ is the uncertainty of reconstruction. Obviously, the bias-variance tradeoff can reveal more detailed information. The term bias$^2$ describes its accuracy (about the deviation from the true values), the variance conveys the precision (about the errors) of the constraint. Theoretically, minimizing risk corresponds to a balance between bias and variance. In cosmology, this promising approach has been widely utilized to find an effective way of obtaining information about the dark energy equation of state $w(z)$ \cite{huterer2002parameterization,zheng2014constraints}. In order to investigate the influence from fiducial model on the risk, we respectively consider the fiducial $\Lambda$CDM model with $\Omega_m=0.305$ and $w$CDM model with $w=-1.027$ in the combination of JLA and complementary probes \cite{betoule2014improved}.

\subsubsection{Accuracy}

\begin{figure}
\includegraphics[width=0.4\textwidth]{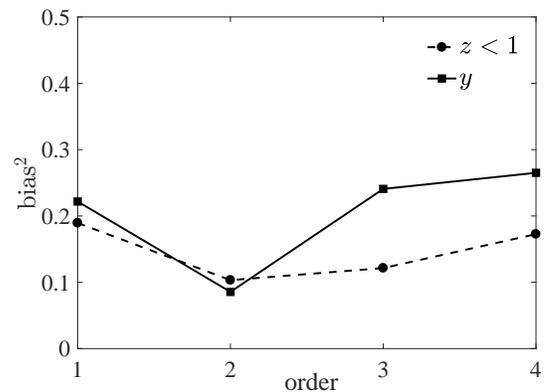}
\caption{ Bias square in the $z$-redshift and $y$-redshift of the JLA compilation with different cosmographic series truncations at the basis of fiducial $\Lambda$CDM model. }  \label{bias figure}
\end{figure}

Accuracy is a deviation from the true value, which can be expressed by the bias square. In Fig. \ref{bias figure} we show the bias$^2$ of current data at the basis of fiducial $\Lambda$CDM model. First, we find that all of the bias are small, which indicates that the cosmographic models fit well with the JLA data. It implies that the cosmography is enough accurate to describe the observational JLA data. This is not difficult to understand, because about 99\% of the JLA data are at low redshift. Thus, application of the JLA data in cosmology would be a very useful strategy. Second, we see that bias square slightly increases for higher order. Last but important, we obtain that the $z$-redshift and $y$-redshift both favor the 2 order, which indicates that expansion up to the jerk term is in the best agreement with the true values.

\subsubsection{Precision}

\begin{figure}
\includegraphics[width=0.4\textwidth]{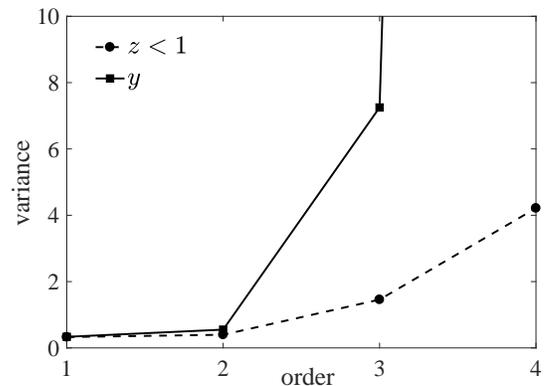}
\caption{ Variance in the $z$-redshift and $y$-redshift of the JLA compilation with different cosmographic series truncations. }  \label{var figure}
\end{figure}

Precision is usually statistic, and represents the errors. Variance, the set of error, is independent of the fiducial model. In Fig. \ref{var figure}, we plot the variance of cosmographic model in $z$-redshift and $y$-redshift. First, variances at low order in these two redshift spaces are both small, almost zero. It indicates that current observational data can present a precise measurement on the parameters $q_0$ and $j_0$. Second, we note that variance at the 3 order starts to increase rapidly, especially for the $y$-redshift, which means that current data cannot give physical measurement on the  $s_0$ term, even higher orders. However, it has enough information for us to infer that the universe will be to continue accelerate or slow down. Third, we should admit that variance in the $y$-redshift space at the 4 order is larger than that of the $z$-redshift.

\subsubsection{Risk}

\begin{figure}
\includegraphics[width=4.2cm,height=4.1cm]{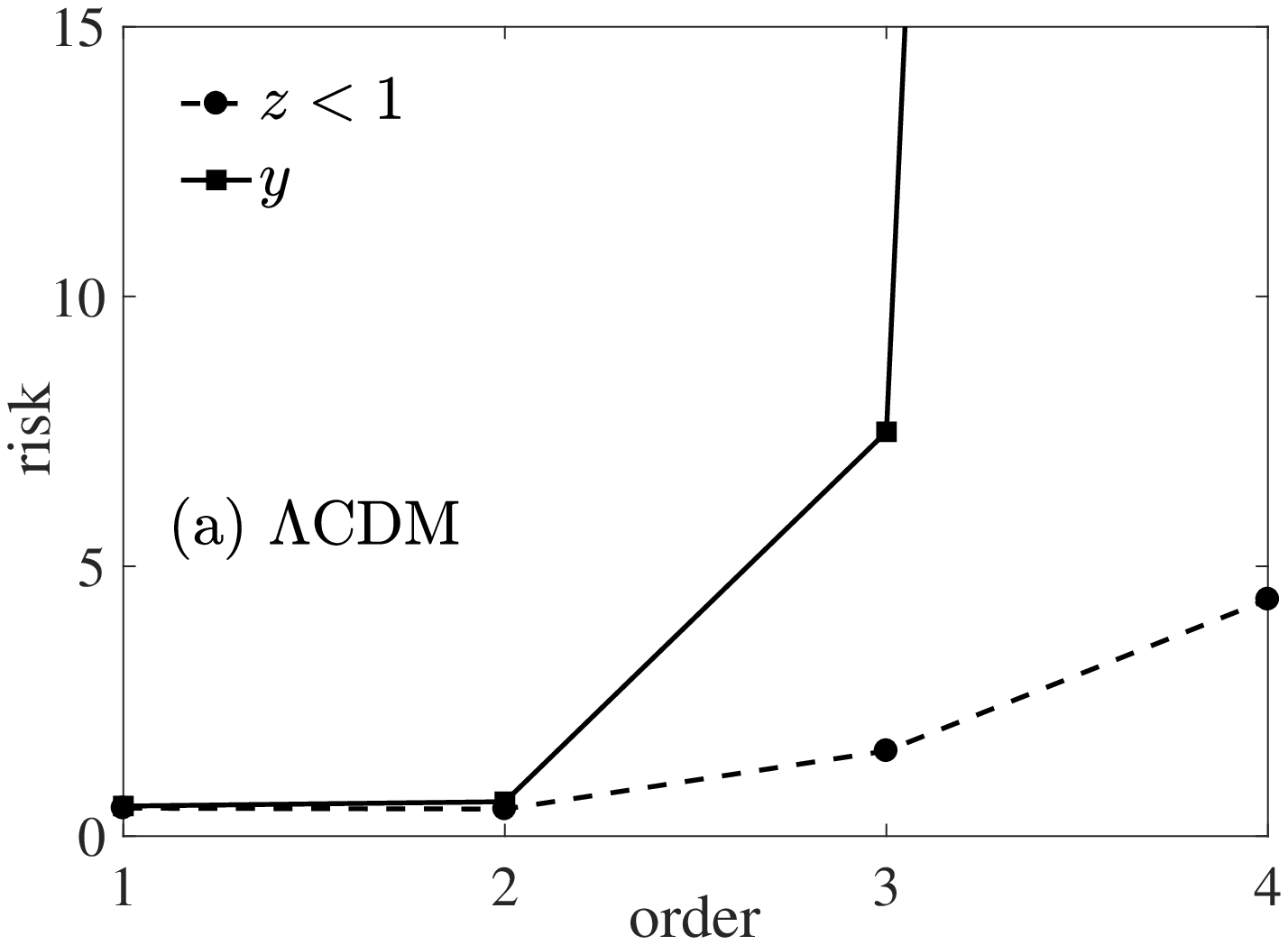}
\includegraphics[width=4.2cm,height=4.1cm]{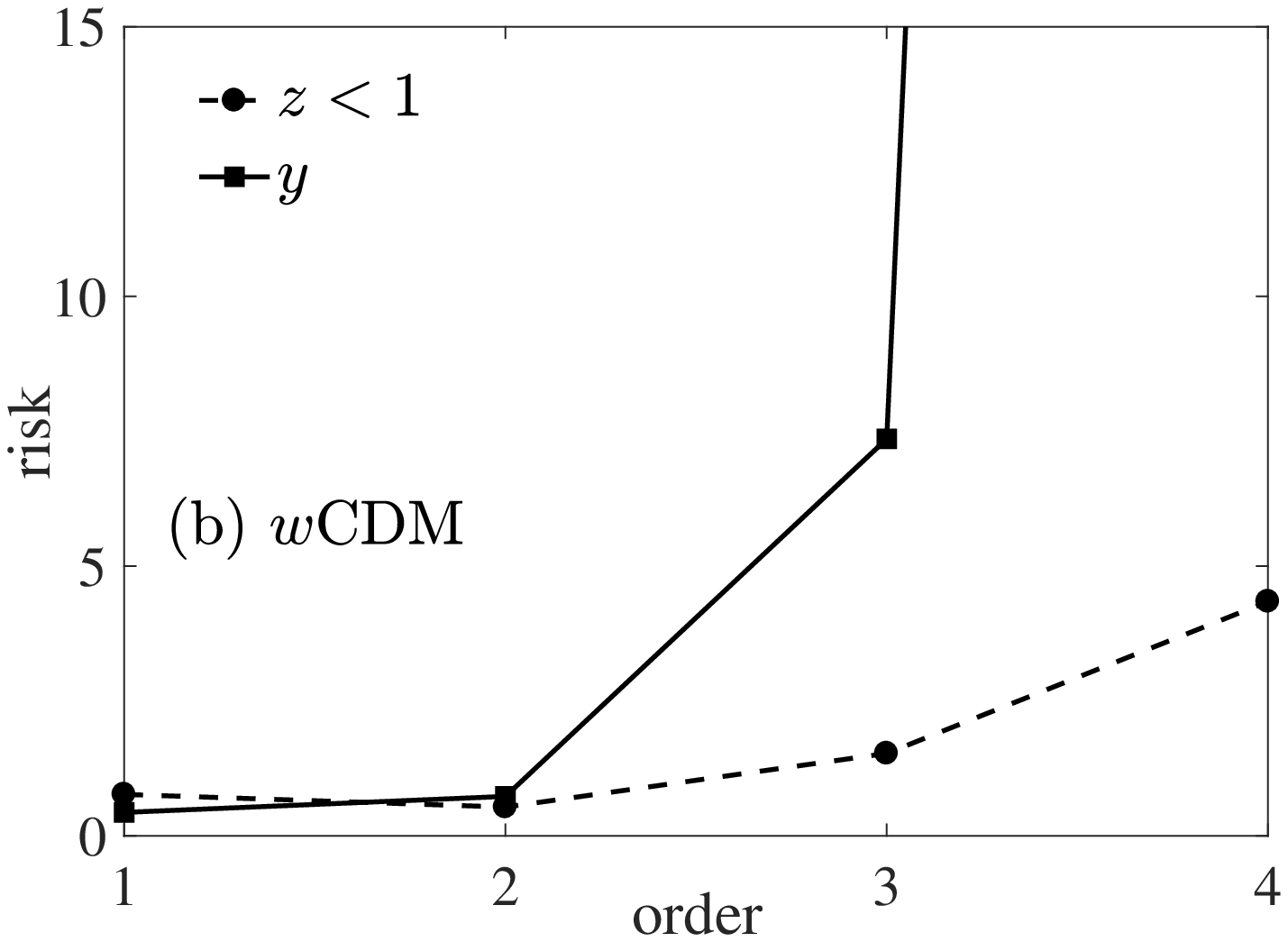}
\caption{ Risk with different cosmographic series truncations in diverse fiducial models. The panel (a) is for the $\Lambda$CDM model, and the panel (b) is for $w$CDM model. }  \label{risk figure}
\end{figure}

Risk is used to balance the bias square and variance, and to find which series truncation is the best description of the observational data. Due to the model-dependence of bias square, in this section we also investigate the influence of different fiducial models on the final risk analysis. In Fig. \ref{risk figure}, we plot the risk for fiducial $\Lambda$CDM model and $w$CDM model. From the comparison between two panels, we first find that risk affected by the fiducial model is so little. They both favor that cosmography up to the $j_0$ term is a better choice to describe current JLA data. This consequence is consistent with our previous work via the \textit{F}-test \cite{xia2012cosmography}. It also proves that the risk analysis is a stable and scientific tool to analyze the convergence problem.

From the bias-variance tradeoff, we conclude that the JLA data is so precise that the cosmographic model at $z$-redshift and $y$-redshift both can present an estimation with high accuracy. Of course, the introduction of $y$-redshift can improve the cosmography study to higher redshift region, even to the early epoch. With its help, past universe can be understood more objective. Meanwhile, the risk analysis is also stable. Effect by the redshift parameter and fiducial model is not significant.

\begin{figure}
\includegraphics[width=0.4\textwidth]{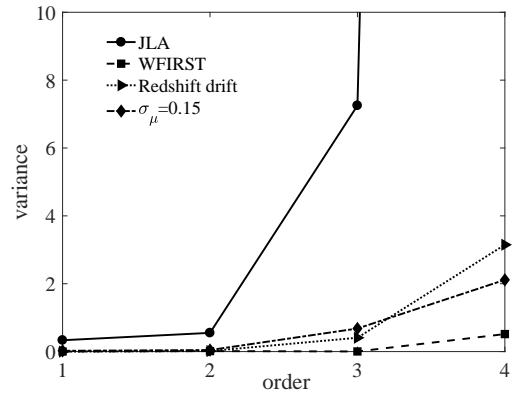}
\caption{ Variance for different cosmographic series truncations in the $y$-redshift by current and future observations.}  \label{future figure}
\end{figure}

\subsection{Forecasting}
\begin{figure*}
\includegraphics[width=5.0cm,height=4.6cm]{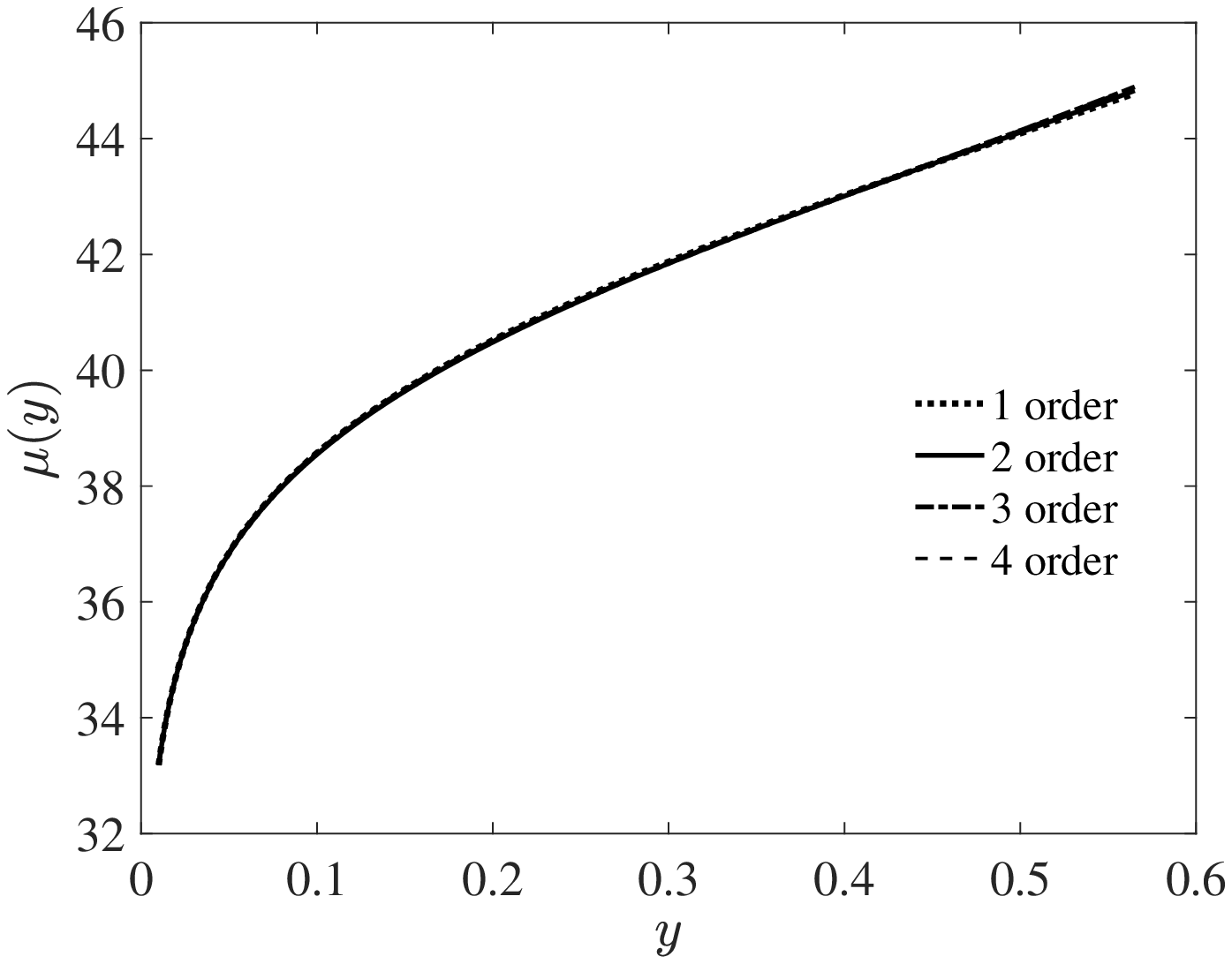}
\includegraphics[width=5.0cm,height=4.6cm]{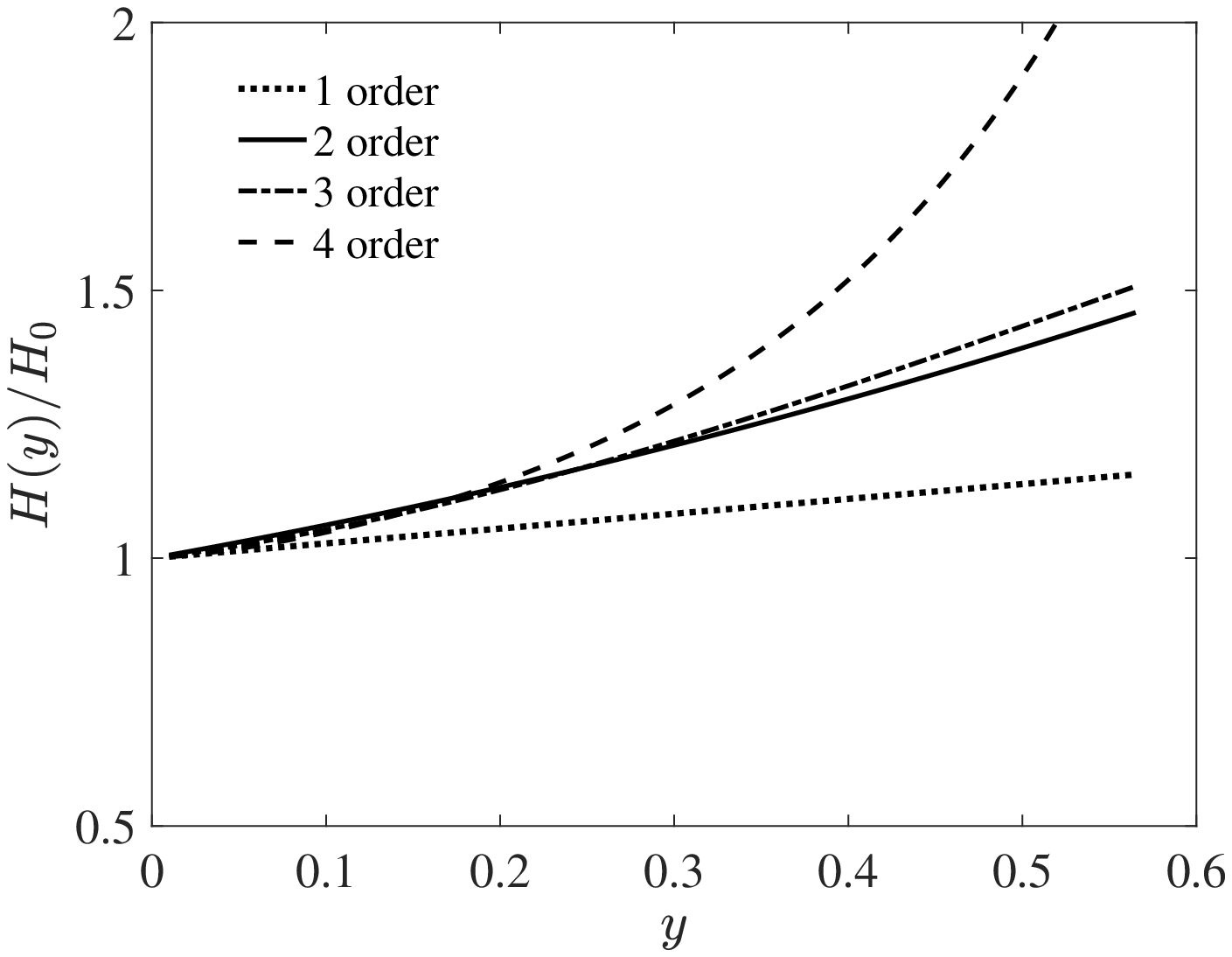}
\includegraphics[width=5.0cm,height=4.6cm]{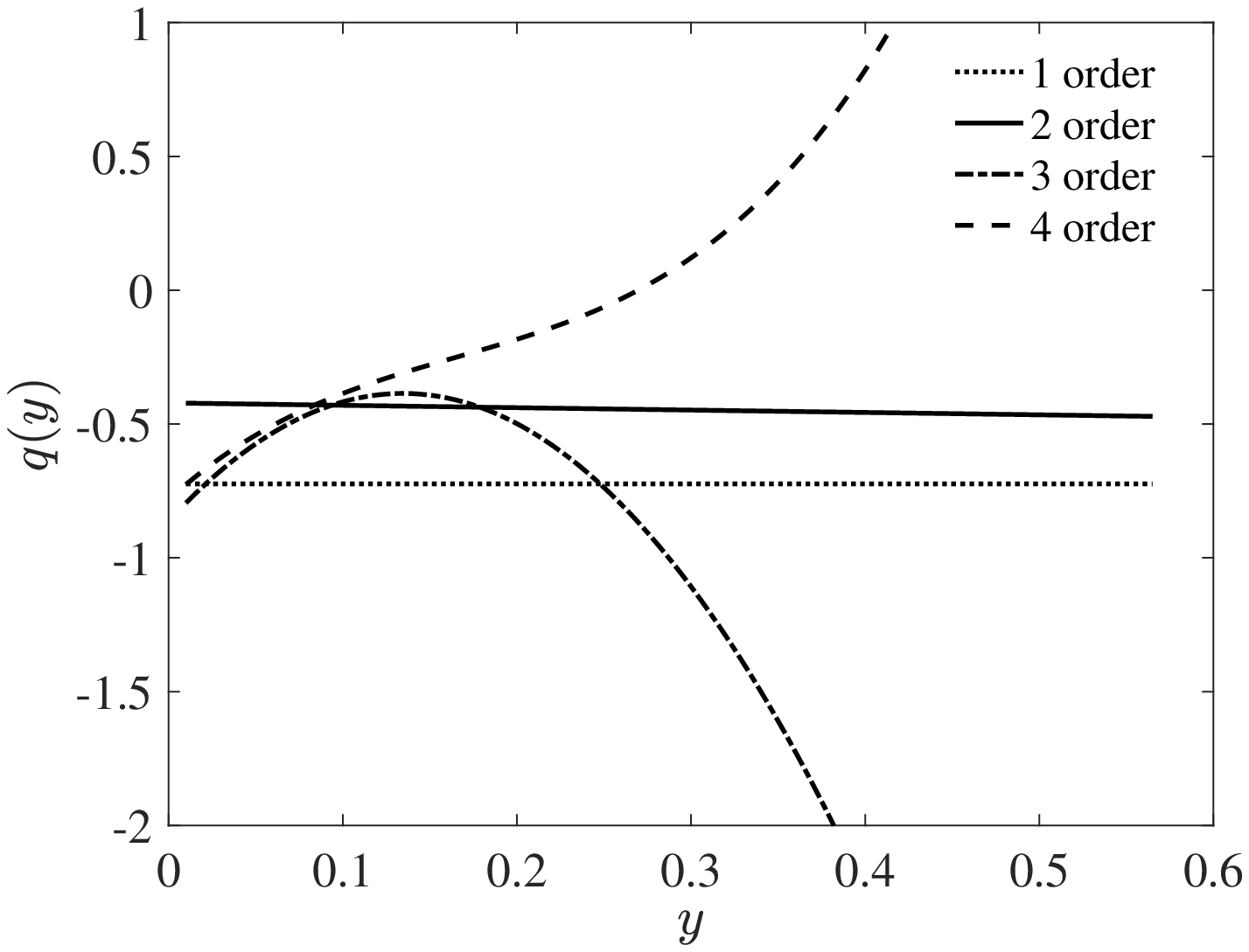}
\caption{Cosmographic distance modulus, Hubble parameter and deceleration parameter with different orders in the $y$-redshift. }  \label{qz figure}
\end{figure*}

Above analysis shows that cosmography at high order suffers unphysical estimation, i.e., large variance. We anticipate that future observation is able to give tighter constraints on the cosmography, with the improvement of observational precision, thus leading to a relaxation of the convergence problem. In this section, we forecast the constraint from future WFIRST and redshift drift on cosmography. In order to test the constraint from mock distance modulus with flat errors $\sigma_{\mu}=0.15$, we also generate some data following the work in Ref. \cite{busti2015cosmography}. Comparison in Table \ref{JLA results} shows that future measurement can improve the constraints. For example, compared with $\sigma_{l_0} = 2478.44$ from the JLA sample, the redshift drift gives a more robust constraint on the parameters, e.g., $\sigma_{l_0} = 1299.17$, almost improving by double than current JLA data. Due to future measurements mainly focus on the high redshift region, we make a comparison on variance for the $y$-redshift in Fig. \ref{future figure}. On the one hand, we find that all the future observations can improve the constraints at low order with high significant. Especially, the redshift drift can present an error $\sigma_{q_0} \sim 10^{-5}$ for the 1 order. On the other hand, the future observations, including the mock data with $\sigma_{\mu}=0.15$ all improve the constraint at high order dramatically. Variances in these cases are much smaller than current data. Thus, we can see that the bias-trade off is effective to estimate the cosmographic problem. The future observations also have the potential to solve the cosmographic problem.

\section{Values of the cosmography}  \label{values}
\begin{figure*}
\includegraphics[width=4.2cm,height=4.2cm]{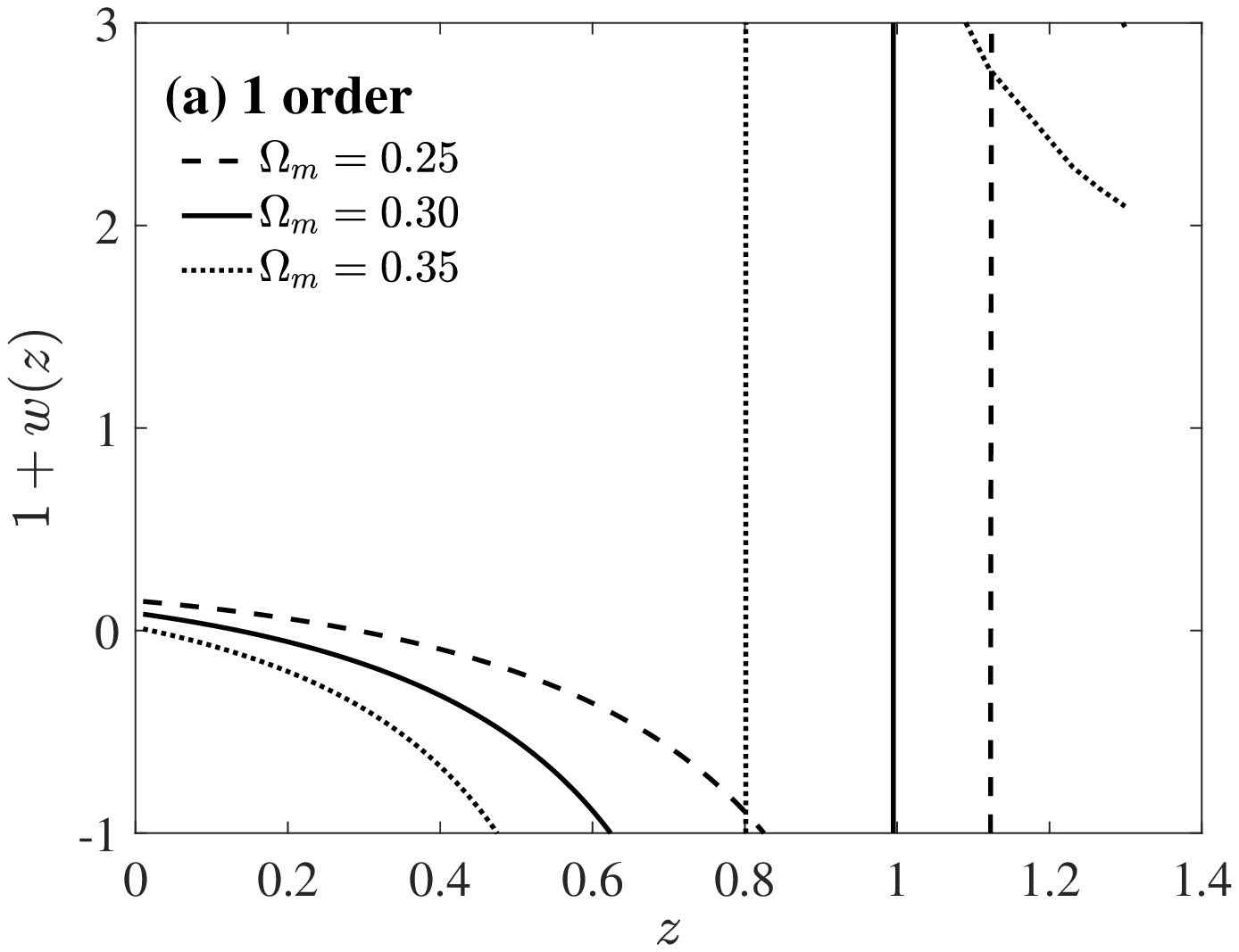}
\includegraphics[width=4.2cm,height=4.2cm]{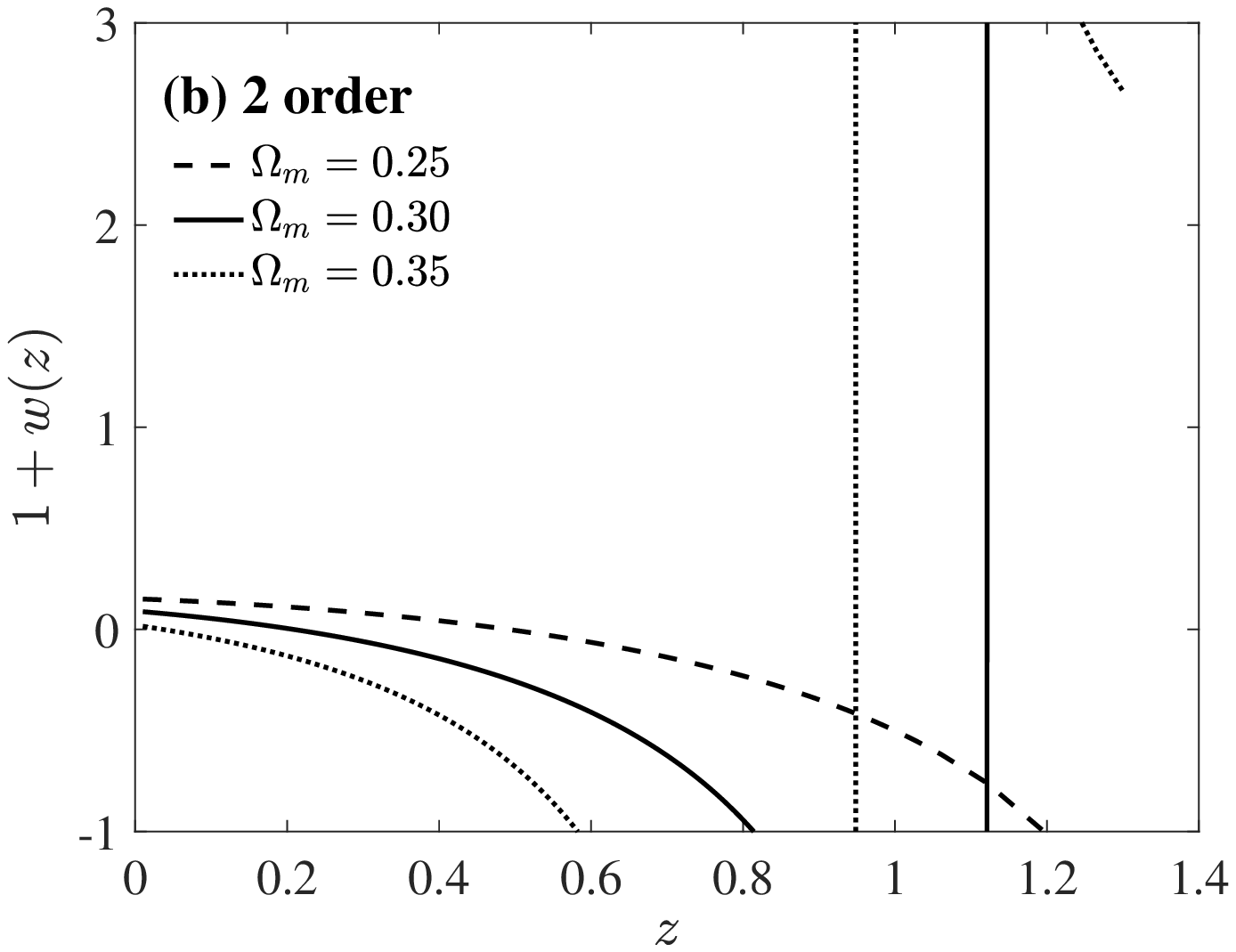}
\includegraphics[width=4.2cm,height=4.2cm]{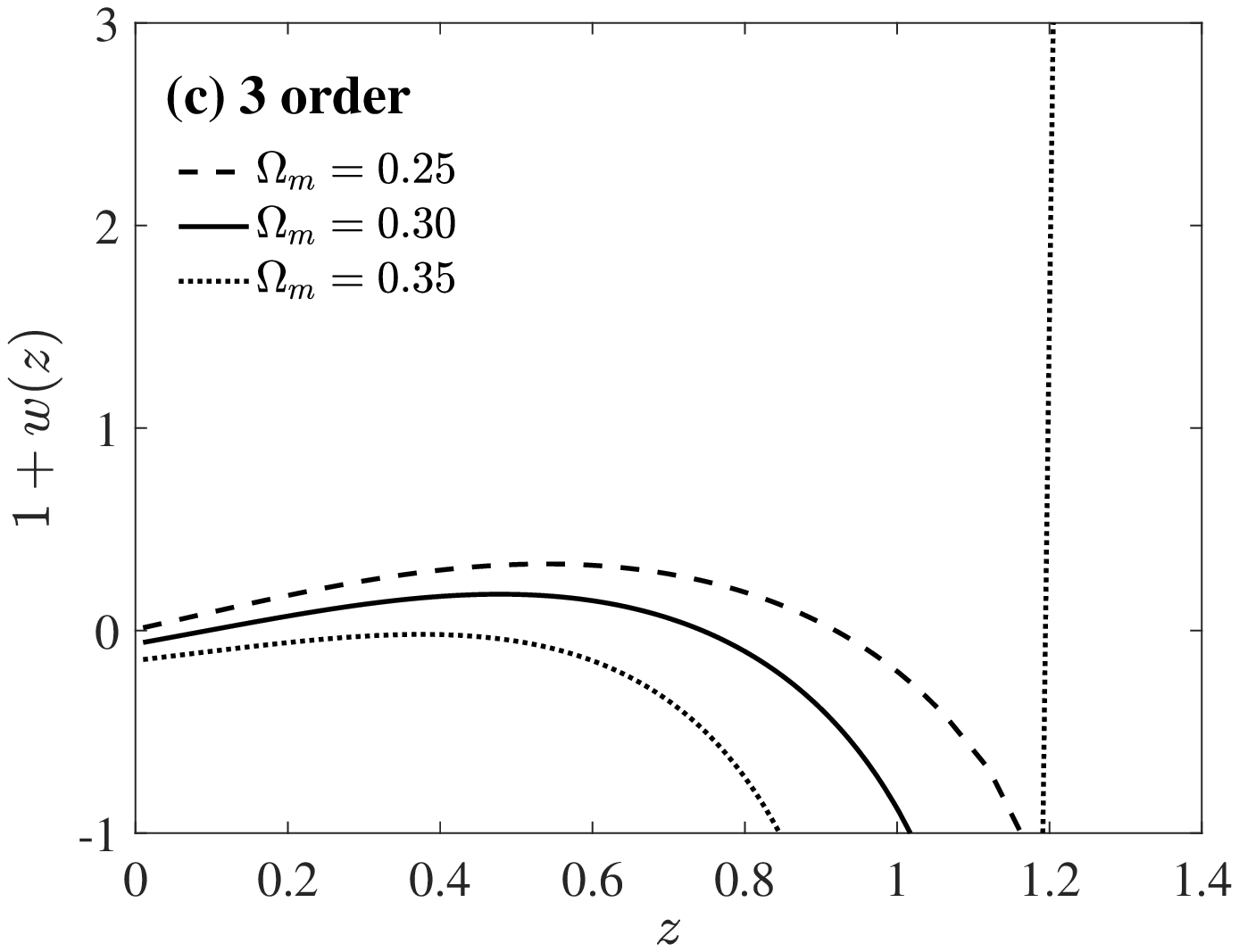}
\includegraphics[width=4.2cm,height=4.2cm]{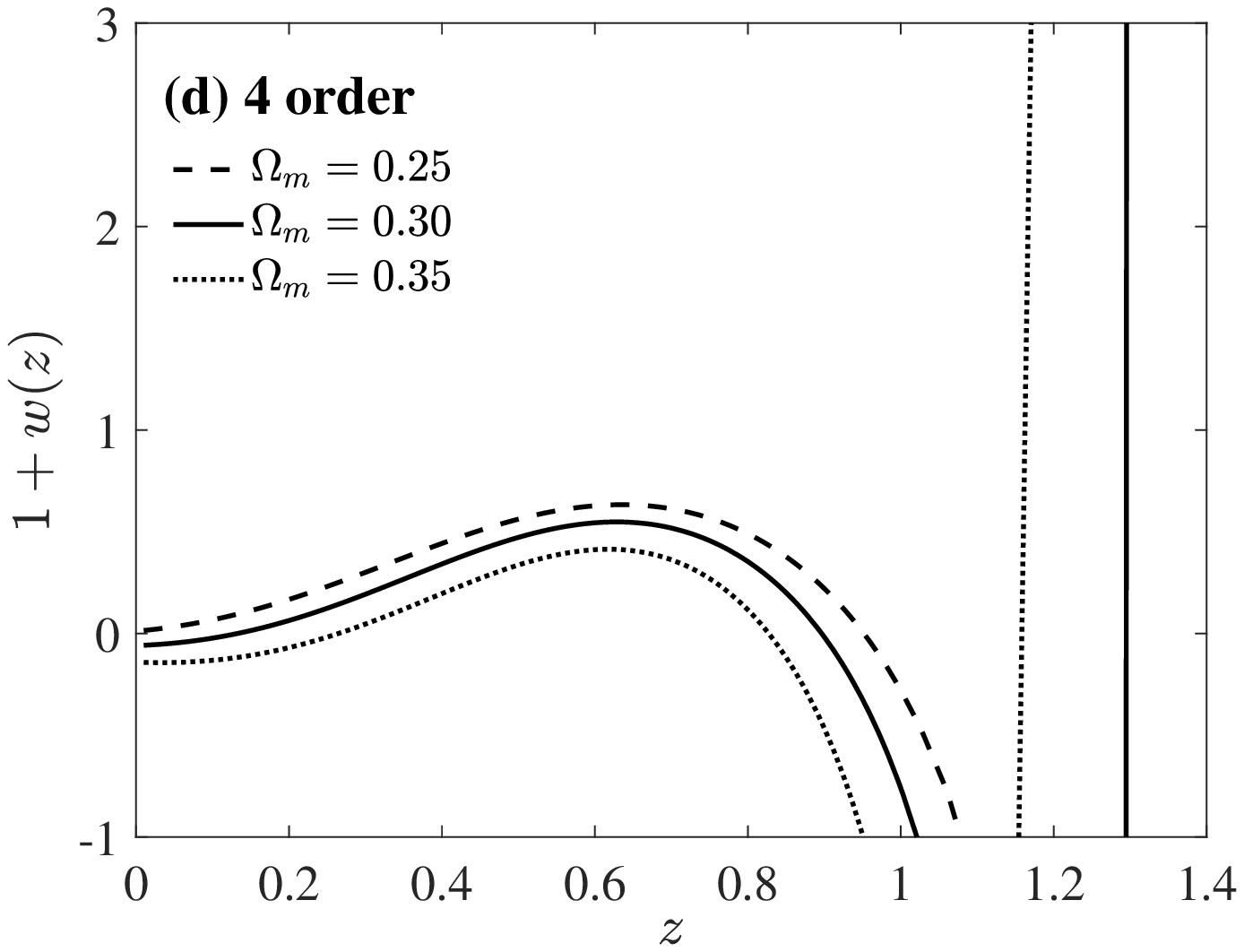}
\caption{Cosmographic EoS for different orders with matter density $\Omega_m=$0.25, 0.30 and 0.35, respectively.  }  \label{w figure}
\end{figure*}
\begin{figure*}
\includegraphics[width=4.2cm,height=4.2cm]{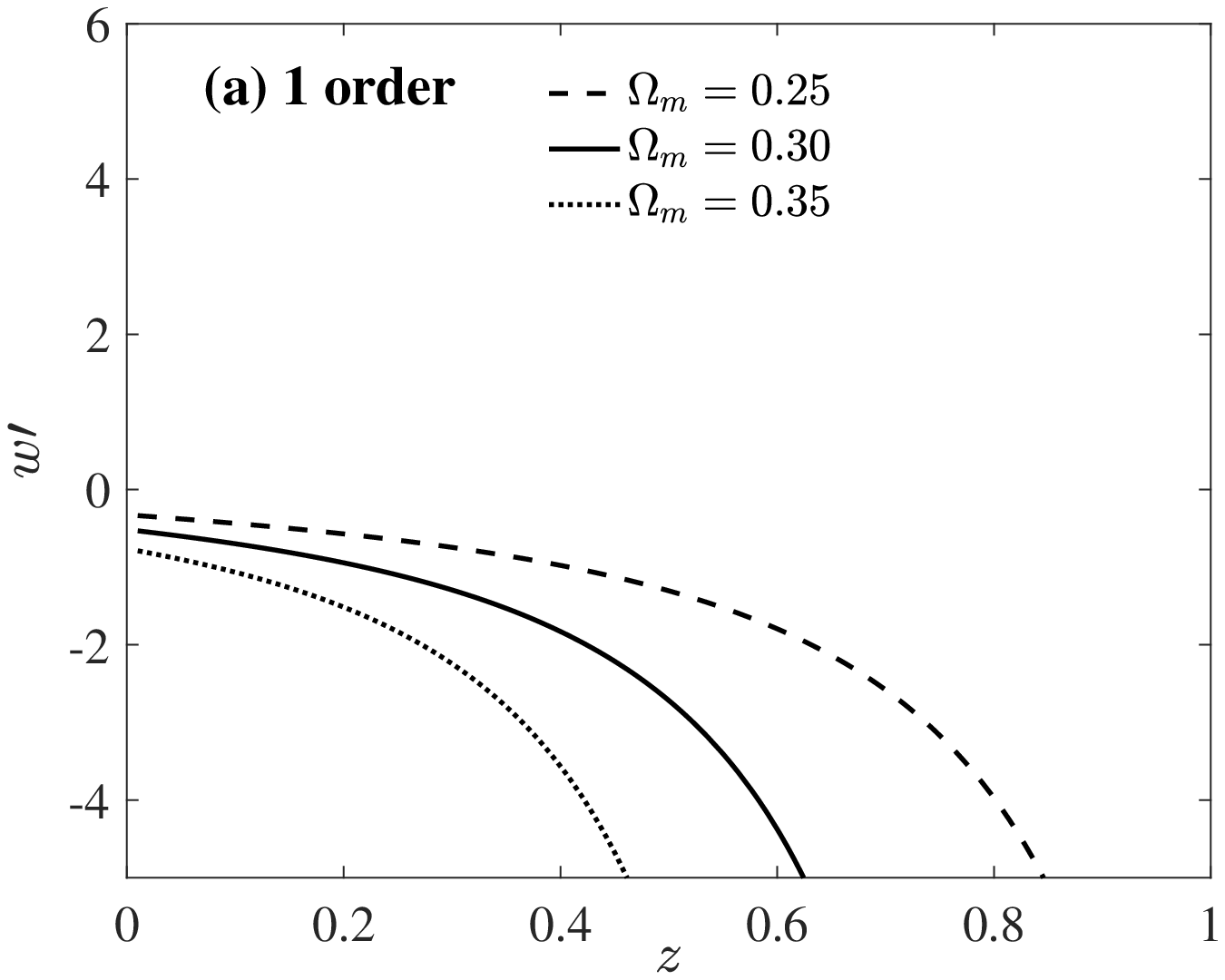}
\includegraphics[width=4.2cm,height=4.2cm]{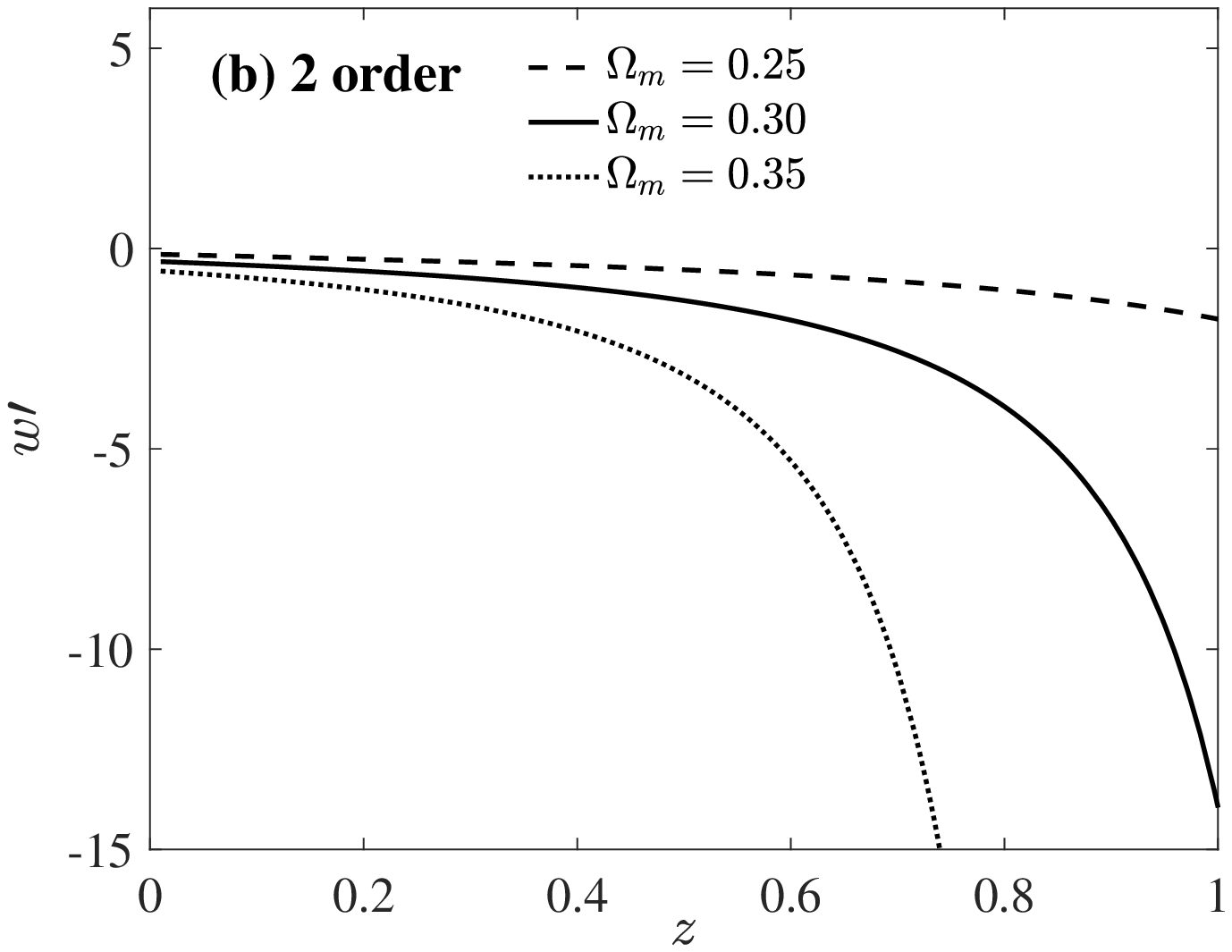}
\includegraphics[width=4.2cm,height=4.2cm]{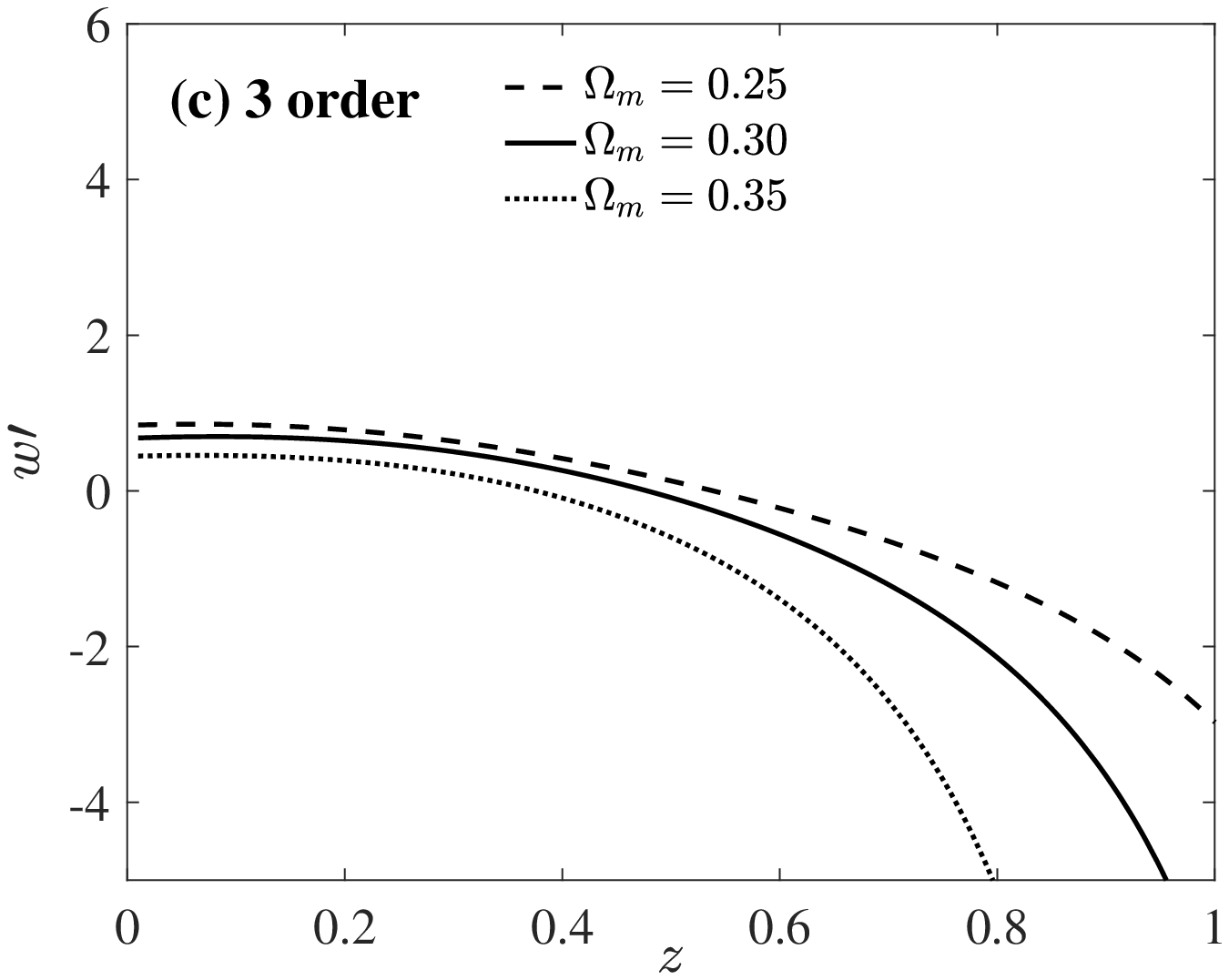}
\includegraphics[width=4.2cm,height=4.2cm]{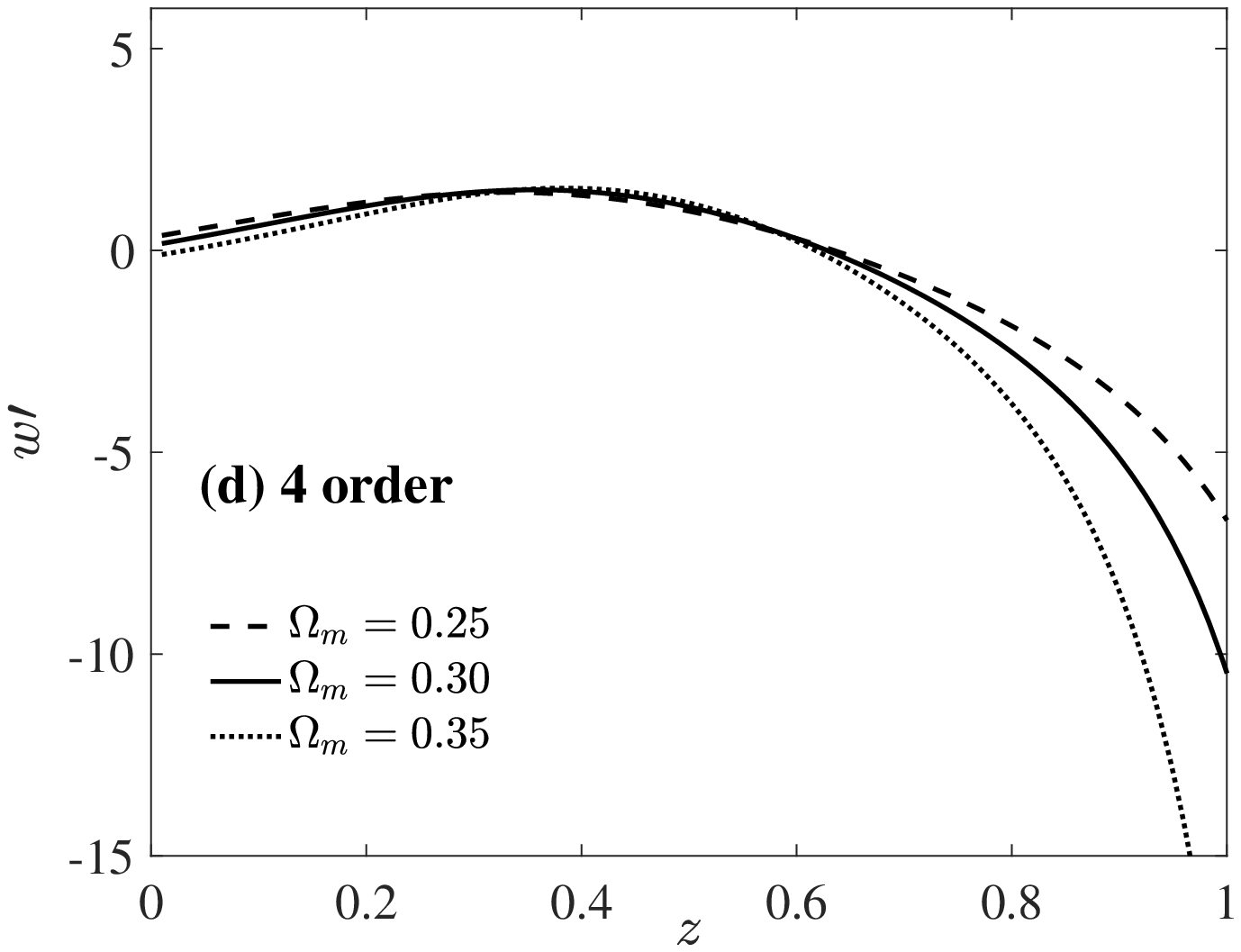}
\caption{Derivative of dark energy EoS $\textmd{d}w/\textmd{d}z$ for different cosmographic orders with varying parameter $\Omega_m$ from 0.25 to 0.35.}  \label{dw figure}
\end{figure*}

In previous work, cosmography has been widely used to reconstruct some special cosmological parameters, because of its model-independence. In this section, we are interested in investigating its values to report what information we can obtain from the cosmography, and what we cannot get.

\subsection{Deceleration parameter}
\label{deceleration factor}

Deceleration parameter is important for its sharp sense on the cosmic expansion. Especially, its negative (positive) sign immediately indicates the accelerating (decelerating) expansion. However, it is not an observable quantity temporarily. Most studies were performed in multiform parameterized $q(z)$. Therefore, a model-independent analysis is appreciated.

In the right panel of Fig. \ref{qz figure}, we plot the reconstructed deceleration factor over $y$-redshift using the best-fit values of supernova data. We find that $q(y)$ in various series truncations are quite different. It strongly depends on the order of Taylor expansion. Therefore, it is difficult to obtain a model-independent and stable estimation on the deceleration parameter via cosmography.

In order to find the reason why we cannot obtain a stable estimation on deceleration parameter, we also compare the cosmographic distance modulus and Hubble parameter for different orders. For the distance modulus, they are almost indistinguishable at all redshift, which indicates that cosmography fit well with the observational data. However, for the Hubble parameter, we can only obtain a relatively stable estimation at low redshift $y < 0.2$. For high redshift, they gradually deviate from each other. When we extract the information of deceleration parameter, we only can obtain a similar estimation at redshift $y \approx 0.1$. This comparison tells us that despite the cosmographic models fit well with the data, their contradiction become more and more prominent, with our increasing requirement on cosmic expansion study. Therefore, it is difficult to obtain a more detailed expansion history via cosmography.

In fact, Fig. \ref{qz figure} implies that a dynamical measurement may be useful to solve this contradiction. In our previous work \cite{xia2012cosmography}, we found that inclusion of Hubble parameter data can lead to stronger constraints on cosmographic parameters. In Ref. \cite{neben2013beyond}, the authors also showed that the distance indicator cannot directly measure $q_0$ with both accuracy and precision. However, the redshift drift may be possible to do it. Therefore, it is reasonable for us to anticipate that inclusion of the dynamical redshift drift could present a much more stable evaluation on cosmic expansion history.

\subsection{Dark energy equation of state}  \label{EoS}

In previous work, cosmography was often used to reconstruct the dynamical cosmological model. For example, with two cosmographic parameters  $(q_{0}, j_{0})$, one can derive the constant equation of state (EoS) dark energy model \cite{demianski2012high}
\begin{eqnarray}
        \Omega_{m}(q_{0}, j_{0}) &=& \frac{2 (j_{0} - q_{0} - 2 q_{0}^{2})}{1 + 2 j_{0}
        - 6 q_{0}}  , \nonumber \\
        w_{0}(q_{0}, j_{0}) &=& \frac{1 + 2 j_{0} - 6 q_{0}}{-3 + 6 q_{0}} .
\end{eqnarray}
However, it needs a background model. In order to get an undamaged map of dark energy, our study is in the normal cosmological model
\be   \label{Hz EoS}
H^2(z) = H_0^2 \left [\Omega_m (1+z)^3 + (1-\Omega_m) \exp \Big[ 3 \int_0^z \frac{1+w(z)}{1+z} \Big] dz \right] .
\ee
In our analysis, we do not impose any style of dark energy, but the common $w(z)$. Solving the Eq. \eqref{Hz EoS}, we obtain
\be  \label{EoS diff}
1 + w(z) = \frac{1}{3} \frac{[H^2(z) - H^2_0 \Omega_m (1+z)^3]' (1+z)}{H^2(z) - H^2_0 \Omega_m (1+z)^3} ,
\ee
where the prime denotes the derivative with respect to redshift $z$. For the Eq. \eqref{EoS diff}, we note that the denominator may be zero for $H(z)^2=H_0^2 \Omega_m (1+z)^3$. This case may leads to a singularity in the EoS reconstruction. In Fig. \ref{w figure}, we plot the reconstruction of dark energy with different cosmographic series. In order to investigate the influence of matter density parameter, we relax parameter $\Omega_m$ from 0.25 to 0.35. On the one hand, we find that cosmography all favor a cosmological constant  EoS at recently. On the other hand, however, a reliable estimation on $w(z)$ is difficult to obtain. In addition, we find that $w(z)$ at redshift $z \sim 0.8$ has a sharp change, independent of the matter density parameter.

Usually, it is difficult to determine the EoS constant or varying model independently. A model-independent analysis of the derivative of EoS can be studied using the cosmography
\be
w' = \frac{\textmd{d} w}{\textmd{d} z} .
\ee
In Fig. \ref{dw figure}, we plot the derivative $w' (z)$ by cosmography with different series. We also consider the parameter $\Omega_m$ in a wide region. At first, we find that $w'(z=0)$ is not zero in different cosmographic models. This indicates that a constant EoS dark energy model may be inappropriate.  Moreover, $w'(z)$ at low redshift are generally negative. Thus, the cosmography may suggest a varying EoS and more complicated dark energy candidate. Linear EoS like  $w(z)=w_0 + w_a z$  may be improper. According to above picture, we infer that cosmography may favor a dark energy with $w(z) = -1 +w_a z + w_b z^2 + \cdots$, where $w_a <0$ and $w_b \neq 0$. However, an accurate determination on the derivative $w'(z)$ need more data to join in, because it also strongly depends on the cosmographic series truncation.

\section{Conclusion and discussion}
\label{conclusion}

In the present paper, we analyse the problem of cosmography using the bias-variance tradeoff, and investigate its values.

To solve the convergence issue in cosmography, an improved redshift $y=z/(1+z)$ was introduced. Using the bias-variance tradeoff, we find that the $y$-redshift produces bigger variances at high orders. For the low cosmographic order (i.e., 1 order and 2 order), $y$-redshift does not bring bigger errors, but a nearly identical variance as $z$-redshift. For the JLA data, we find that most of them distribute in low redshift region with high precision. So that, the $z$-redshift is enough accurate to describe the data. Although the $y$-redshift does not present a smaller bias than $z$-redshift for the JLA data, it still can ensure the correctness of cosmography at high redshift.

Minimizing risk, it suggests that expansion up to the $j_0$ term is the best choice for current supernova data, regardless of the $z$-redshift or $y$-redshift. We also test the influence of fiducial model on risk analysis. The comparison demonstrates that it is trivial. Although previous \textit{F}-test also obtained a similar result, our paper is not a repeated work using more data. Our analysis finds out deeper reason of the convergence issue. First, it not only can tell us the convergence problem is in the accuracy or the precision, but also can provide us more objective information about how serious the divergence problem is. Because if the crux lies on the accuracy, the convergence problem maybe still cannot be solved even thought more data were included. Second, in previous work, most focus were on the pursuit of ``sweet spot", which has masked the physical meaning of $y$-redshift. In our study, we not only find it is influenced by the distribution of data, but also forecast whether future observations can solve the convergence problem. Our analysis in Fig. \ref{qz figure} and Section \ref{deceleration factor} also indicates that the dynamical measurement is a potential clue to solve this problem.

Our forecast finds that future WFIRST and redshift drift can significantly improve the constraints. Therefore, inclusion of dynamical measurement such as Hubble parameter data, redshift drift, etc. may be able to improve the constraint in accuracy and precision with high significance. As studied in our previous work \cite{xia2012cosmography}, inclusion of the $H(z)$ data can lead to stronger constraints on the cosmographic parameters. This discovery is helpful to understand or solve the convergence issue of supernova data. This is because dynamical probe like the canonical redshift drift can provide direct measurement to the cosmic expansion history. While distance measurement is geometric. As studied in Ref.  \cite{maor2001limitations}, the luminosity distance determines the EoS $w$ through a multiple integral relation that smears out much information. For the redshift drift, it not only directly measures the change of Hubble parameter, but also can be realized via multiple wavebands and methods \cite{loeb1998direct,darling2012toward}. Moreover, it is immune from extra systematic errors, and does not need photometric calibration etc. Recently, a test in German Vacuum Tower Telescope demonstrates that the Laser Frequency Combs also have an advantage with long-term calibration precision, accuracy to realize the redshift drift experiment \cite{steinmetz2008laser}.

Our investigation also promotes the study of the values of cosmography. In previous work, most attention were focus on a special model. However, our analysis presents a almost undamaged map of dark energy. It breaks the limitation of extrapolation to other models. Setting the dark energy $w(z)$ as free, we find that cosmography cannot give reliable estimations on $q(z)$ and $w(z)$. However, we find that it does not favor a constant EoS, but a complicated $w(z)$, such as $w(z) = -1 +w_a z + w_b z^2 + \cdots$, where $w_a <0$ and $w_b \neq 0$. These estimations are useful for modelling the dark energy.

Cosmography has been an useful tool with great potential to study the cosmology. For the dark energy, it was usually reconstructed by parametrization, such as the Chevallier-Polarski-Linder \cite{chevallier2001accelerating,linder2003exploring},  Jassal-Bagla-Padmanabhan \cite{2005MNRAS.356L..11J}; or the non-parameterization, such as the Gaussian processes \cite{seikel2012reconstruction,nair2014exploring}, principal component analysis \cite{crittenden2012fables,zheng2014constraints}. Cosmography is another model-independent method to assess dark energy models. Moreover, cosmography has also been widely used in another fields, such as to test the power of supernova data \cite{ma2016statistical}. Therefore, we have to say that cosmography is an important method to study the cosmology. Our study provides a straightforward and scientific reference. Of course, we will also devote ourselves to improving the cosmography study in our future work. We would like to study the influence of the inclusion of BAO and CMB data on cosmography. Throughout previous work, we find that many different observational data or combinations favor the best cosmography with 2 order. In our future work, we are also interested in exploring their subtle relations to further understand the cosmography. Moreover, we also has an interest to improve the cosmographic problem by proposing some other physical redshift.

\section*{Acknowledgments}

We quite appreciate the anonymous referee for the suggestions to improve this manuscript. We thank Wei Zheng, Si-Yu Li, Yang Liu and Yong-Ping Li for their help on the calculation. J.-Q. Xia is supported by the National Youth Thousand Talents Program and the National Science Foundation of China under grant No. 11422323. H. Li is supported in part by NSFC under grant Nos. 11033005 and 11322325 and by the 973 program under
grant No. 2010CB83300. M.-J. Zhang is funded by China Postdoctoral Science Foundation under grant No. 2015M581173. The research is also supported by the Strategic Priority Research Program ¡°The Emergence of Cosmological Structures¡± of the Chinese Academy of Sciences, grant No. XDB09000000.

\bibliography{cosmography}

\begin{thebibliography}{71}
\expandafter\ifx\csname natexlab\endcsname\relax\def\natexlab#1{#1}\fi
\expandafter\ifx\csname bibnamefont\endcsname\relax
  \def\bibnamefont#1{#1}\fi
\expandafter\ifx\csname bibfnamefont\endcsname\relax
  \def\bibfnamefont#1{#1}\fi
\expandafter\ifx\csname citenamefont\endcsname\relax
  \def\citenamefont#1{#1}\fi
\expandafter\ifx\csname url\endcsname\relax
  \def\url#1{\texttt{#1}}\fi
\expandafter\ifx\csname urlprefix\endcsname\relax\def\urlprefix{URL }\fi
\providecommand{\bibinfo}[2]{#2}
\providecommand{\eprint}[2][]{\url{#2}}

\bibitem[{\citenamefont{Carroll et~al.}(1992)\citenamefont{Carroll, Press, and
  Turner}}]{carroll1992cosmological}
\bibinfo{author}{\bibfnamefont{S.~M.} \bibnamefont{Carroll}},
  \bibinfo{author}{\bibfnamefont{W.~H.} \bibnamefont{Press}}, \bibnamefont{and}
  \bibinfo{author}{\bibfnamefont{E.~L.} \bibnamefont{Turner}},
  \bibinfo{journal}{Annu. Rev. Astron. Astrophys.}
  \textbf{\bibinfo{volume}{30}}, \bibinfo{pages}{499} (\bibinfo{year}{1992}).

\bibitem[{\citenamefont{Caldwell et~al.}(2003)\citenamefont{Caldwell,
  Kamionkowski, and Weinberg}}]{caldwell2003phantom}
\bibinfo{author}{\bibfnamefont{R.~R.} \bibnamefont{Caldwell}},
  \bibinfo{author}{\bibfnamefont{M.}~\bibnamefont{Kamionkowski}},
  \bibnamefont{and} \bibinfo{author}{\bibfnamefont{N.~N.}
  \bibnamefont{Weinberg}}, \bibinfo{journal}{Phys. Rev. Lett.}
  \textbf{\bibinfo{volume}{91}}, \bibinfo{pages}{071301}
  (\bibinfo{year}{2003}).

\bibitem[{\citenamefont{Feng et~al.}(2005)\citenamefont{Feng, Wang, and
  Zhang}}]{feng2005wang}
\bibinfo{author}{\bibfnamefont{B.}~\bibnamefont{Feng}},
  \bibinfo{author}{\bibfnamefont{X.}~\bibnamefont{Wang}}, \bibnamefont{and}
  \bibinfo{author}{\bibfnamefont{X.}~\bibnamefont{Zhang}},
  \bibinfo{journal}{Phys. Lett., B} \textbf{\bibinfo{volume}{607}},
  \bibinfo{pages}{35} (\bibinfo{year}{2005}).

\bibitem[{\citenamefont{Barrow and Cotsakis}(1988)}]{barrow1988inflation}
\bibinfo{author}{\bibfnamefont{J.~D.} \bibnamefont{Barrow}} \bibnamefont{and}
  \bibinfo{author}{\bibfnamefont{S.}~\bibnamefont{Cotsakis}},
  \bibinfo{journal}{Phys. Lett. B} \textbf{\bibinfo{volume}{214}},
  \bibinfo{pages}{515} (\bibinfo{year}{1988}).

\bibitem[{\citenamefont{Dvali et~al.}(2000)\citenamefont{Dvali, Gabadadze, and
  Porrati}}]{dvali20004d}
\bibinfo{author}{\bibfnamefont{G.}~\bibnamefont{Dvali}},
  \bibinfo{author}{\bibfnamefont{G.}~\bibnamefont{Gabadadze}},
  \bibnamefont{and} \bibinfo{author}{\bibfnamefont{M.}~\bibnamefont{Porrati}},
  \bibinfo{journal}{Phys. Lett. B} \textbf{\bibinfo{volume}{485}},
  \bibinfo{pages}{208} (\bibinfo{year}{2000}).

\bibitem[{\citenamefont{Lema{\^\i}tre}(1933)}]{lemaitre1933univers}
\bibinfo{author}{\bibfnamefont{G.}~\bibnamefont{Lema{\^\i}tre}},
  \bibinfo{journal}{Annales de la Societe Scietifique de Bruxelles}
  \textbf{\bibinfo{volume}{53}}, \bibinfo{pages}{51} (\bibinfo{year}{1933}).

\bibitem[{\citenamefont{Tolman}(1934)}]{tolman1934effect}
\bibinfo{author}{\bibfnamefont{R.~C.} \bibnamefont{Tolman}},
  \bibinfo{journal}{Proceedings of the national academy of sciences of the
  United States of America} \textbf{\bibinfo{volume}{20}}, \bibinfo{pages}{169}
  (\bibinfo{year}{1934}).

\bibitem[{\citenamefont{Bondi}(1947)}]{bondi1947spherically}
\bibinfo{author}{\bibfnamefont{H.}~\bibnamefont{Bondi}}, \bibinfo{journal}{Mon.
  Not. R. Astron. Soc.} \textbf{\bibinfo{volume}{107}}, \bibinfo{pages}{410}
  (\bibinfo{year}{1947}).

\bibitem[{\citenamefont{Chiba and Nakamura}(1998)}]{chiba1998luminosity}
\bibinfo{author}{\bibfnamefont{T.}~\bibnamefont{Chiba}} \bibnamefont{and}
  \bibinfo{author}{\bibfnamefont{T.}~\bibnamefont{Nakamura}},
  \bibinfo{journal}{Progress of theoretical physics}
  \textbf{\bibinfo{volume}{100}}, \bibinfo{pages}{1077} (\bibinfo{year}{1998}).

\bibitem[{\citenamefont{Visser}(2004)}]{visser2004jerk}
\bibinfo{author}{\bibfnamefont{M.}~\bibnamefont{Visser}},
  \bibinfo{journal}{Classical and Quantum Gravity}
  \textbf{\bibinfo{volume}{21}}, \bibinfo{pages}{2603} (\bibinfo{year}{2004}).

\bibitem[{\citenamefont{Catto{\"e}n and Visser}(2007)}]{cattoen2007hubble}
\bibinfo{author}{\bibfnamefont{C.}~\bibnamefont{Catto{\"e}n}} \bibnamefont{and}
  \bibinfo{author}{\bibfnamefont{M.}~\bibnamefont{Visser}},
  \bibinfo{journal}{Classical and Quantum Gravity}
  \textbf{\bibinfo{volume}{24}}, \bibinfo{pages}{5985} (\bibinfo{year}{2007}).

\bibitem[{\citenamefont{Aviles et~al.}(2012)\citenamefont{Aviles, Gruber,
  Luongo, and Quevedo}}]{aviles2012cosmography}
\bibinfo{author}{\bibfnamefont{A.}~\bibnamefont{Aviles}},
  \bibinfo{author}{\bibfnamefont{C.}~\bibnamefont{Gruber}},
  \bibinfo{author}{\bibfnamefont{O.}~\bibnamefont{Luongo}}, \bibnamefont{and}
  \bibinfo{author}{\bibfnamefont{H.}~\bibnamefont{Quevedo}},
  \bibinfo{journal}{Physical Review D} \textbf{\bibinfo{volume}{86}},
  \bibinfo{pages}{123516} (\bibinfo{year}{2012}).

\bibitem[{\citenamefont{Catto{\"e}n and
  Visser}(2008)}]{cattoen2008cosmographic}
\bibinfo{author}{\bibfnamefont{C.}~\bibnamefont{Catto{\"e}n}} \bibnamefont{and}
  \bibinfo{author}{\bibfnamefont{M.}~\bibnamefont{Visser}},
  \bibinfo{journal}{Phys. Rev. D} \textbf{\bibinfo{volume}{78}},
  \bibinfo{pages}{063501} (\bibinfo{year}{2008}).

\bibitem[{\citenamefont{Capozziello et~al.}(2011)\citenamefont{Capozziello,
  Lazkoz, and Salzano}}]{capozziello2011comprehensive}
\bibinfo{author}{\bibfnamefont{S.}~\bibnamefont{Capozziello}},
  \bibinfo{author}{\bibfnamefont{R.}~\bibnamefont{Lazkoz}}, \bibnamefont{and}
  \bibinfo{author}{\bibfnamefont{V.}~\bibnamefont{Salzano}},
  \bibinfo{journal}{Physical Review D} \textbf{\bibinfo{volume}{84}},
  \bibinfo{pages}{124061} (\bibinfo{year}{2011}).

\bibitem[{\citenamefont{Xia et~al.}(2012)\citenamefont{Xia, Vitagliano,
  Liberati, and Viel}}]{xia2012cosmography}
\bibinfo{author}{\bibfnamefont{J.-Q.} \bibnamefont{Xia}},
  \bibinfo{author}{\bibfnamefont{V.}~\bibnamefont{Vitagliano}},
  \bibinfo{author}{\bibfnamefont{S.}~\bibnamefont{Liberati}}, \bibnamefont{and}
  \bibinfo{author}{\bibfnamefont{M.}~\bibnamefont{Viel}},
  \bibinfo{journal}{Phys. Rev. D} \textbf{\bibinfo{volume}{85}},
  \bibinfo{pages}{043520} (\bibinfo{year}{2012}).

\bibitem[{\citenamefont{Vitagliano et~al.}(2010)\citenamefont{Vitagliano, Xia,
  Liberati, and Viel}}]{vitagliano2010high}
\bibinfo{author}{\bibfnamefont{V.}~\bibnamefont{Vitagliano}},
  \bibinfo{author}{\bibfnamefont{J.-Q.} \bibnamefont{Xia}},
  \bibinfo{author}{\bibfnamefont{S.}~\bibnamefont{Liberati}}, \bibnamefont{and}
  \bibinfo{author}{\bibfnamefont{M.}~\bibnamefont{Viel}},
  \bibinfo{journal}{Journal of Cosmology and Astroparticle Physics}
  \textbf{\bibinfo{volume}{2010}}, \bibinfo{pages}{005} (\bibinfo{year}{2010}).

\bibitem[{\citenamefont{Lazkoz et~al.}(2013)\citenamefont{Lazkoz, Alcaniz,
  Escamilla-Rivera, Salzano, and Sendra}}]{lazkoz2013bao}
\bibinfo{author}{\bibfnamefont{R.}~\bibnamefont{Lazkoz}},
  \bibinfo{author}{\bibfnamefont{J.}~\bibnamefont{Alcaniz}},
  \bibinfo{author}{\bibfnamefont{C.}~\bibnamefont{Escamilla-Rivera}},
  \bibinfo{author}{\bibfnamefont{V.}~\bibnamefont{Salzano}}, \bibnamefont{and}
  \bibinfo{author}{\bibfnamefont{I.}~\bibnamefont{Sendra}},
  \bibinfo{journal}{Journal of Cosmology and Astroparticle Physics}
  \textbf{\bibinfo{volume}{12}}, \bibinfo{pages}{005} (\bibinfo{year}{2013}).

\bibitem[{\citenamefont{Busti et~al.}(2015)\citenamefont{Busti, de~la
  Cruz-Dombriz, Dunsby, and S{\'a}ez-G{\'o}mez}}]{busti2015cosmography}
\bibinfo{author}{\bibfnamefont{V.~C.} \bibnamefont{Busti}},
  \bibinfo{author}{\bibfnamefont{{\'A}.}~\bibnamefont{de~la Cruz-Dombriz}},
  \bibinfo{author}{\bibfnamefont{P.~K.~S.} \bibnamefont{Dunsby}},
  \bibnamefont{and}
  \bibinfo{author}{\bibfnamefont{D.}~\bibnamefont{S{\'a}ez-G{\'o}mez}},
  \bibinfo{journal}{Physical Review D} \textbf{\bibinfo{volume}{92}},
  \bibinfo{pages}{123512} (\bibinfo{year}{2015}).

\bibitem[{\citenamefont{Suzuki et~al.}(2012)\citenamefont{Suzuki, Rubin, Lidman
  et~al.}}]{suzuki2012hubble}
\bibinfo{author}{\bibfnamefont{N.}~\bibnamefont{Suzuki}},
  \bibinfo{author}{\bibfnamefont{D.}~\bibnamefont{Rubin}},
  \bibinfo{author}{\bibfnamefont{C.}~\bibnamefont{Lidman}},
  \bibnamefont{et~al.}, \bibinfo{journal}{Astrophys. J.}
  \textbf{\bibinfo{volume}{746}}, \bibinfo{pages}{85} (\bibinfo{year}{2012}).

\bibitem[{\citenamefont{Sandage}(1962)}]{sandage1962change}
\bibinfo{author}{\bibfnamefont{A.}~\bibnamefont{Sandage}},
  \bibinfo{journal}{Astrophys. J.} \textbf{\bibinfo{volume}{136}},
  \bibinfo{pages}{319} (\bibinfo{year}{1962}).

\bibitem[{\citenamefont{Liske et~al.}(2008{\natexlab{a}})\citenamefont{Liske,
  Grazian, Vanzella et~al.}}]{liske2008cosmic}
\bibinfo{author}{\bibfnamefont{J.}~\bibnamefont{Liske}},
  \bibinfo{author}{\bibfnamefont{A.}~\bibnamefont{Grazian}},
  \bibinfo{author}{\bibfnamefont{E.}~\bibnamefont{Vanzella}},
  \bibnamefont{et~al.}, \bibinfo{journal}{Mon. Not. R. Astron. Soc.}
  \textbf{\bibinfo{volume}{386}}, \bibinfo{pages}{1192}
  (\bibinfo{year}{2008}{\natexlab{a}}).

\bibitem[{\citenamefont{Liske et~al.}(2008{\natexlab{b}})\citenamefont{Liske,
  Grazian, Vanzella et~al.}}]{liske2008elt}
\bibinfo{author}{\bibfnamefont{J.}~\bibnamefont{Liske}},
  \bibinfo{author}{\bibfnamefont{A.}~\bibnamefont{Grazian}},
  \bibinfo{author}{\bibfnamefont{E.}~\bibnamefont{Vanzella}},
  \bibnamefont{et~al.}, \bibinfo{journal}{The Messenger}
  \textbf{\bibinfo{volume}{133}}, \bibinfo{pages}{10}
  (\bibinfo{year}{2008}{\natexlab{b}}).

\bibitem[{\citenamefont{Corasaniti et~al.}(2007)\citenamefont{Corasaniti,
  Huterer, and Melchiorri}}]{corasaniti2007exploring}
\bibinfo{author}{\bibfnamefont{P.-S.} \bibnamefont{Corasaniti}},
  \bibinfo{author}{\bibfnamefont{D.}~\bibnamefont{Huterer}}, \bibnamefont{and}
  \bibinfo{author}{\bibfnamefont{A.}~\bibnamefont{Melchiorri}},
  \bibinfo{journal}{Physical Review D} \textbf{\bibinfo{volume}{75}},
  \bibinfo{pages}{062001} (\bibinfo{year}{2007}).

\bibitem[{\citenamefont{Zhang and Liu}(2014)}]{zhang2014observational}
\bibinfo{author}{\bibfnamefont{M.-J.} \bibnamefont{Zhang}} \bibnamefont{and}
  \bibinfo{author}{\bibfnamefont{W.-B.} \bibnamefont{Liu}},
  \bibinfo{journal}{The European Physical Journal C}
  \textbf{\bibinfo{volume}{74}}, \bibinfo{pages}{1} (\bibinfo{year}{2014}).

\bibitem[{\citenamefont{Geng et~al.}(2015)\citenamefont{Geng, Li, Zhang, and
  Zhang}}]{geng2015redshift}
\bibinfo{author}{\bibfnamefont{J.-J.} \bibnamefont{Geng}},
  \bibinfo{author}{\bibfnamefont{Y.-H.} \bibnamefont{Li}},
  \bibinfo{author}{\bibfnamefont{J.-F.} \bibnamefont{Zhang}}, \bibnamefont{and}
  \bibinfo{author}{\bibfnamefont{X.}~\bibnamefont{Zhang}},
  \bibinfo{journal}{The European Physical Journal C}
  \textbf{\bibinfo{volume}{75}}, \bibinfo{pages}{1} (\bibinfo{year}{2015}).

\bibitem[{\citenamefont{Li et~al.}(2013)\citenamefont{Li, Liao, Wu, Yu, and
  Zhu}}]{li2013probing}
\bibinfo{author}{\bibfnamefont{Z.}~\bibnamefont{Li}},
  \bibinfo{author}{\bibfnamefont{K.}~\bibnamefont{Liao}},
  \bibinfo{author}{\bibfnamefont{P.}~\bibnamefont{Wu}},
  \bibinfo{author}{\bibfnamefont{H.}~\bibnamefont{Yu}}, \bibnamefont{and}
  \bibinfo{author}{\bibfnamefont{Z.-H.} \bibnamefont{Zhu}},
  \bibinfo{journal}{Phys. Rev. D} \textbf{\bibinfo{volume}{88}},
  \bibinfo{pages}{023003} (\bibinfo{year}{2013}).

\bibitem[{\citenamefont{Uzan et~al.}(2008)\citenamefont{Uzan, Clarkson, and
  Ellis}}]{uzan2008time}
\bibinfo{author}{\bibfnamefont{J.-P.} \bibnamefont{Uzan}},
  \bibinfo{author}{\bibfnamefont{C.}~\bibnamefont{Clarkson}}, \bibnamefont{and}
  \bibinfo{author}{\bibfnamefont{G.~F.~R.} \bibnamefont{Ellis}},
  \bibinfo{journal}{Physical Review Letters} \textbf{\bibinfo{volume}{100}},
  \bibinfo{pages}{191303} (\bibinfo{year}{2008}).

\bibitem[{\citenamefont{Yu et~al.}(2014)\citenamefont{Yu, Zhang, and
  Pen}}]{yu2014method}
\bibinfo{author}{\bibfnamefont{H.-R.} \bibnamefont{Yu}},
  \bibinfo{author}{\bibfnamefont{T.-J.} \bibnamefont{Zhang}}, \bibnamefont{and}
  \bibinfo{author}{\bibfnamefont{U.-L.} \bibnamefont{Pen}},
  \bibinfo{journal}{Physical Review Letters} \textbf{\bibinfo{volume}{113}},
  \bibinfo{pages}{041303} (\bibinfo{year}{2014}).

\bibitem[{\citenamefont{Sahni et~al.}(2003)\citenamefont{Sahni, Saini,
  Starobinsky, and Alam}}]{sahni2003statefinder}
\bibinfo{author}{\bibfnamefont{V.}~\bibnamefont{Sahni}},
  \bibinfo{author}{\bibfnamefont{T.~D.} \bibnamefont{Saini}},
  \bibinfo{author}{\bibfnamefont{A.~A.} \bibnamefont{Starobinsky}},
  \bibnamefont{and} \bibinfo{author}{\bibfnamefont{U.}~\bibnamefont{Alam}},
  \bibinfo{journal}{Journal of Experimental and Theoretical Physics Letters}
  \textbf{\bibinfo{volume}{77}}, \bibinfo{pages}{201} (\bibinfo{year}{2003}).

\bibitem[{\citenamefont{Alam et~al.}(2003)\citenamefont{Alam, Sahni, Saini, and
  Starobinsky}}]{alam2003exploring}
\bibinfo{author}{\bibfnamefont{U.}~\bibnamefont{Alam}},
  \bibinfo{author}{\bibfnamefont{V.}~\bibnamefont{Sahni}},
  \bibinfo{author}{\bibfnamefont{T.~D.} \bibnamefont{Saini}}, \bibnamefont{and}
  \bibinfo{author}{\bibfnamefont{A.}~\bibnamefont{Starobinsky}},
  \bibinfo{journal}{Monthly Notices of the Royal Astronomical Society}
  \textbf{\bibinfo{volume}{344}}, \bibinfo{pages}{1057} (\bibinfo{year}{2003}).

\bibitem[{\citenamefont{Jimenez and Loeb}(2008)}]{jimenez2008constraining}
\bibinfo{author}{\bibfnamefont{R.}~\bibnamefont{Jimenez}} \bibnamefont{and}
  \bibinfo{author}{\bibfnamefont{A.}~\bibnamefont{Loeb}}, \bibinfo{journal}{The
  Astrophysical Journal} \textbf{\bibinfo{volume}{573}}, \bibinfo{pages}{37}
  (\bibinfo{year}{2008}).

\bibitem[{\citenamefont{Simon et~al.}(2005)\citenamefont{Simon, Verde, and
  Jimenez}}]{simon2005constraints}
\bibinfo{author}{\bibfnamefont{J.}~\bibnamefont{Simon}},
  \bibinfo{author}{\bibfnamefont{L.}~\bibnamefont{Verde}}, \bibnamefont{and}
  \bibinfo{author}{\bibfnamefont{R.}~\bibnamefont{Jimenez}},
  \bibinfo{journal}{Phys. Rev. D} \textbf{\bibinfo{volume}{71}},
  \bibinfo{pages}{123001} (\bibinfo{year}{2005}).

\bibitem[{\citenamefont{Stern et~al.}(2010)\citenamefont{Stern, Jimenez, Verde,
  Kamionkowski, and Stanford}}]{stern2010cosmic}
\bibinfo{author}{\bibfnamefont{D.}~\bibnamefont{Stern}},
  \bibinfo{author}{\bibfnamefont{R.}~\bibnamefont{Jimenez}},
  \bibinfo{author}{\bibfnamefont{L.}~\bibnamefont{Verde}},
  \bibinfo{author}{\bibfnamefont{M.}~\bibnamefont{Kamionkowski}},
  \bibnamefont{and} \bibinfo{author}{\bibfnamefont{S.~A.}
  \bibnamefont{Stanford}}, \bibinfo{journal}{Journal of Cosmology and
  Astroparticle Physics} \textbf{\bibinfo{volume}{2010}}, \bibinfo{pages}{008}
  (\bibinfo{year}{2010}).

\bibitem[{\citenamefont{Gaztanaga et~al.}(2009)\citenamefont{Gaztanaga, Cabre,
  and Hui}}]{gaztanaga2009clustering}
\bibinfo{author}{\bibfnamefont{E.}~\bibnamefont{Gaztanaga}},
  \bibinfo{author}{\bibfnamefont{A.}~\bibnamefont{Cabre}}, \bibnamefont{and}
  \bibinfo{author}{\bibfnamefont{L.}~\bibnamefont{Hui}},
  \bibinfo{journal}{Monthly Notices of the Royal Astronomical Society}
  \textbf{\bibinfo{volume}{399}}, \bibinfo{pages}{1663} (\bibinfo{year}{2009}).

\bibitem[{\citenamefont{Moresco et~al.}(2012)\citenamefont{Moresco, Cimatti,
  Jimenez et~al.}}]{moresco2012improved}
\bibinfo{author}{\bibfnamefont{M.}~\bibnamefont{Moresco}},
  \bibinfo{author}{\bibfnamefont{A.}~\bibnamefont{Cimatti}},
  \bibinfo{author}{\bibfnamefont{R.}~\bibnamefont{Jimenez}},
  \bibnamefont{et~al.}, \bibinfo{journal}{Journal of Cosmology and
  Astroparticle Physics} \textbf{\bibinfo{volume}{08}}, \bibinfo{pages}{006}
  (\bibinfo{year}{2012}).

\bibitem[{\citenamefont{Busca et~al.}(2013)\citenamefont{Busca, Delubac, Rich
  et~al.}}]{delubac2013baryon}
\bibinfo{author}{\bibfnamefont{N.~G.} \bibnamefont{Busca}},
  \bibinfo{author}{\bibfnamefont{T.}~\bibnamefont{Delubac}},
  \bibinfo{author}{\bibfnamefont{J.}~\bibnamefont{Rich}}, \bibnamefont{et~al.},
  \bibinfo{journal}{Astronomy and Astrophysics} \textbf{\bibinfo{volume}{552}},
  \bibinfo{pages}{A96} (\bibinfo{year}{2013}), \eprint{1211.2616}.

\bibitem[{\citenamefont{Loeb}(1998)}]{loeb1998direct}
\bibinfo{author}{\bibfnamefont{A.}~\bibnamefont{Loeb}},
  \bibinfo{journal}{Astrophys. J. Letters} \textbf{\bibinfo{volume}{499}},
  \bibinfo{pages}{L111} (\bibinfo{year}{1998}).

\bibitem[{\citenamefont{Martinelli et~al.}(2012)\citenamefont{Martinelli,
  Pandolfi, Martins, and Vielzeuf}}]{martinelli2012probing}
\bibinfo{author}{\bibfnamefont{M.}~\bibnamefont{Martinelli}},
  \bibinfo{author}{\bibfnamefont{S.}~\bibnamefont{Pandolfi}},
  \bibinfo{author}{\bibfnamefont{C.~J. A.~P.} \bibnamefont{Martins}},
  \bibnamefont{and} \bibinfo{author}{\bibfnamefont{P.~E.}
  \bibnamefont{Vielzeuf}}, \bibinfo{journal}{Phys. Rev. D}
  \textbf{\bibinfo{volume}{86}}, \bibinfo{pages}{123001}
  (\bibinfo{year}{2012}).

\bibitem[{\citenamefont{Moraes and Polarski}(2011)}]{moraes2011complementarity}
\bibinfo{author}{\bibfnamefont{B.}~\bibnamefont{Moraes}} \bibnamefont{and}
  \bibinfo{author}{\bibfnamefont{D.}~\bibnamefont{Polarski}},
  \bibinfo{journal}{Phys. Rev. D} \textbf{\bibinfo{volume}{84}},
  \bibinfo{pages}{104003} (\bibinfo{year}{2011}).

\bibitem[{\citenamefont{Zhang et~al.}(2014)\citenamefont{Zhang, Qi, and
  Liu}}]{zhang2014cosmic}
\bibinfo{author}{\bibfnamefont{M.-J.} \bibnamefont{Zhang}},
  \bibinfo{author}{\bibfnamefont{J.-Z.} \bibnamefont{Qi}}, \bibnamefont{and}
  \bibinfo{author}{\bibfnamefont{W.-B.} \bibnamefont{Liu}},
  \bibinfo{journal}{International Journal of Theoretical Physics}
  \textbf{\bibinfo{volume}{54}}, \bibinfo{pages}{2456} (\bibinfo{year}{2014}).

\bibitem[{\citenamefont{Balcerzak and Dabrowski}(2014)}]{balcerzak2014redshift}
\bibinfo{author}{\bibfnamefont{A.}~\bibnamefont{Balcerzak}} \bibnamefont{and}
  \bibinfo{author}{\bibfnamefont{M.~P.} \bibnamefont{Dabrowski}},
  \bibinfo{journal}{Physics Letters B} \textbf{\bibinfo{volume}{728}},
  \bibinfo{pages}{15} (\bibinfo{year}{2014}).

\bibitem[{\citenamefont{Lewis and Bridle}(2002)}]{lewis2002cosmological}
\bibinfo{author}{\bibfnamefont{A.}~\bibnamefont{Lewis}} \bibnamefont{and}
  \bibinfo{author}{\bibfnamefont{S.}~\bibnamefont{Bridle}},
  \bibinfo{journal}{Physical Review D} \textbf{\bibinfo{volume}{66}},
  \bibinfo{pages}{103511} (\bibinfo{year}{2002}).

\bibitem[{\citenamefont{Betoule et~al.}(2014)\citenamefont{Betoule, Kessler,
  Guy et~al.}}]{betoule2014improved}
\bibinfo{author}{\bibfnamefont{M.}~\bibnamefont{Betoule}},
  \bibinfo{author}{\bibfnamefont{R.}~\bibnamefont{Kessler}},
  \bibinfo{author}{\bibfnamefont{J.}~\bibnamefont{Guy}}, \bibnamefont{et~al.},
  \bibinfo{journal}{Astronomy \& Astrophysics} \textbf{\bibinfo{volume}{568}},
  \bibinfo{pages}{A22} (\bibinfo{year}{2014}).

\bibitem[{\citenamefont{{Bochner} et~al.}(2015)\citenamefont{{Bochner},
  {Pappas}, and {Dong}}}]{2015ApJ...814....7B}
\bibinfo{author}{\bibfnamefont{B.}~\bibnamefont{{Bochner}}},
  \bibinfo{author}{\bibfnamefont{D.}~\bibnamefont{{Pappas}}}, \bibnamefont{and}
  \bibinfo{author}{\bibfnamefont{M.}~\bibnamefont{{Dong}}},
  \bibinfo{journal}{The Astrophysical Journal} \textbf{\bibinfo{volume}{814}},
  \bibinfo{eid}{7} (\bibinfo{year}{2015}), \eprint{1308.6050}.

\bibitem[{\citenamefont{Spergel et~al.}(2015)\citenamefont{Spergel, Gehrels,
  Baltay, Bennett, Breckinridge, Donahue, Dressler, Gaudi, Greene, Guyon
  et~al.}}]{spergel2015wide}
\bibinfo{author}{\bibfnamefont{D.}~\bibnamefont{Spergel}},
  \bibinfo{author}{\bibfnamefont{N.}~\bibnamefont{Gehrels}},
  \bibinfo{author}{\bibfnamefont{C.}~\bibnamefont{Baltay}},
  \bibinfo{author}{\bibfnamefont{D.}~\bibnamefont{Bennett}},
  \bibinfo{author}{\bibfnamefont{J.}~\bibnamefont{Breckinridge}},
  \bibinfo{author}{\bibfnamefont{M.}~\bibnamefont{Donahue}},
  \bibinfo{author}{\bibfnamefont{A.}~\bibnamefont{Dressler}},
  \bibinfo{author}{\bibfnamefont{B.}~\bibnamefont{Gaudi}},
  \bibinfo{author}{\bibfnamefont{T.}~\bibnamefont{Greene}},
  \bibinfo{author}{\bibfnamefont{O.}~\bibnamefont{Guyon}},
  \bibnamefont{et~al.}, \bibinfo{journal}{arXiv:1503.03757}
  (\bibinfo{year}{2015}).

\bibitem[{\citenamefont{Pasquini et~al.}(2006)\citenamefont{Pasquini,
  Cristiani, Dekker et~al.}}]{pasquini2006codex}
\bibinfo{author}{\bibfnamefont{L.}~\bibnamefont{Pasquini}},
  \bibinfo{author}{\bibfnamefont{S.}~\bibnamefont{Cristiani}},
  \bibinfo{author}{\bibfnamefont{H.}~\bibnamefont{Dekker}},
  \bibnamefont{et~al.}, \bibinfo{journal}{Scientific Requirements for Extremely
  Large Telescopes} \textbf{\bibinfo{volume}{232}}, \bibinfo{pages}{193}
  (\bibinfo{year}{2006}).

\bibitem[{\citenamefont{Pasquini et~al.}(2005)\citenamefont{Pasquini,
  Cristiani, Dekker et~al.}}]{pasquini2005codex}
\bibinfo{author}{\bibfnamefont{L.}~\bibnamefont{Pasquini}},
  \bibinfo{author}{\bibfnamefont{S.}~\bibnamefont{Cristiani}},
  \bibinfo{author}{\bibfnamefont{H.}~\bibnamefont{Dekker}},
  \bibnamefont{et~al.}, \bibinfo{journal}{The Messenger}
  \textbf{\bibinfo{volume}{122}}, \bibinfo{pages}{10} (\bibinfo{year}{2005}).

\bibitem[{\citenamefont{Shafieloo et~al.}(2009)\citenamefont{Shafieloo, Sahni,
  and Starobinsky}}]{shafieloo2009cosmic}
\bibinfo{author}{\bibfnamefont{A.}~\bibnamefont{Shafieloo}},
  \bibinfo{author}{\bibfnamefont{V.}~\bibnamefont{Sahni}}, \bibnamefont{and}
  \bibinfo{author}{\bibfnamefont{A.~A.} \bibnamefont{Starobinsky}},
  \bibinfo{journal}{Phys. Rev. D} \textbf{\bibinfo{volume}{80}},
  \bibinfo{pages}{101301} (\bibinfo{year}{2009}).

\bibitem[{\citenamefont{C{\'a}rdenas et~al.}(2013)\citenamefont{C{\'a}rdenas,
  Bernal, and Bonilla}}]{cardenas2013cosmic}
\bibinfo{author}{\bibfnamefont{V.~H.} \bibnamefont{C{\'a}rdenas}},
  \bibinfo{author}{\bibfnamefont{C.}~\bibnamefont{Bernal}}, \bibnamefont{and}
  \bibinfo{author}{\bibfnamefont{A.}~\bibnamefont{Bonilla}},
  \bibinfo{journal}{Mon. Not. R. Astron. Soc.} \textbf{\bibinfo{volume}{433}},
  \bibinfo{pages}{3534} (\bibinfo{year}{2013}).

\bibitem[{\citenamefont{Li et~al.}(2011)\citenamefont{Li, Wu, and
  Yu}}]{li2011examining}
\bibinfo{author}{\bibfnamefont{Z.}~\bibnamefont{Li}},
  \bibinfo{author}{\bibfnamefont{P.}~\bibnamefont{Wu}}, \bibnamefont{and}
  \bibinfo{author}{\bibfnamefont{H.}~\bibnamefont{Yu}}, \bibinfo{journal}{Phys.
  Lett. B} \textbf{\bibinfo{volume}{695}}, \bibinfo{pages}{1}
  (\bibinfo{year}{2011}).

\bibitem[{\citenamefont{{Lin} et~al.}(2013)\citenamefont{{Lin}, {Wu}, and
  {Yu}}}]{2013PhRvD..87d3502L}
\bibinfo{author}{\bibfnamefont{J.}~\bibnamefont{{Lin}}},
  \bibinfo{author}{\bibfnamefont{P.}~\bibnamefont{{Wu}}}, \bibnamefont{and}
  \bibinfo{author}{\bibfnamefont{H.}~\bibnamefont{{Yu}}},
  \bibinfo{journal}{Phys. Rev. D} \textbf{\bibinfo{volume}{87}},
  \bibinfo{pages}{043502} (\bibinfo{year}{2013}).

\bibitem[{\citenamefont{Maga{\~n}a et~al.}(2014)\citenamefont{Maga{\~n}a,
  C{\'a}rdenas, and Motta}}]{magana2014cosmic}
\bibinfo{author}{\bibfnamefont{J.}~\bibnamefont{Maga{\~n}a}},
  \bibinfo{author}{\bibfnamefont{V.~H.} \bibnamefont{C{\'a}rdenas}},
  \bibnamefont{and} \bibinfo{author}{\bibfnamefont{V.}~\bibnamefont{Motta}},
  \bibinfo{journal}{Journal of Cosmology and Astroparticle Physics}
  \textbf{\bibinfo{volume}{10}}, \bibinfo{pages}{017} (\bibinfo{year}{2014}).

\bibitem[{\citenamefont{{Wang} et~al.}(2016)\citenamefont{{Wang}, {Hu}, {Li},
  and {Li}}}]{2016ApJ...821...60W}
\bibinfo{author}{\bibfnamefont{S.}~\bibnamefont{{Wang}}},
  \bibinfo{author}{\bibfnamefont{Y.}~\bibnamefont{{Hu}}},
  \bibinfo{author}{\bibfnamefont{M.}~\bibnamefont{{Li}}}, \bibnamefont{and}
  \bibinfo{author}{\bibfnamefont{N.}~\bibnamefont{{Li}}},
  \bibinfo{journal}{Astrophys. J.} \textbf{\bibinfo{volume}{821}},
  \bibinfo{pages}{60} (\bibinfo{year}{2016}), \eprint{1509.03461}.

\bibitem[{\citenamefont{{Maga{\~n}a} et~al.}(2017)\citenamefont{{Maga{\~n}a},
  {Motta}, {C{\'a}rdenas}, and {Fo{\"e}x}}}]{2017MNRAS.469...47M}
\bibinfo{author}{\bibfnamefont{J.}~\bibnamefont{{Maga{\~n}a}}},
  \bibinfo{author}{\bibfnamefont{V.}~\bibnamefont{{Motta}}},
  \bibinfo{author}{\bibfnamefont{V.~H.} \bibnamefont{{C{\'a}rdenas}}},
  \bibnamefont{and}
  \bibinfo{author}{\bibfnamefont{G.}~\bibnamefont{{Fo{\"e}x}}},
  \bibinfo{journal}{Mon. Not. R. Astron. Soc.} \textbf{\bibinfo{volume}{469}},
  \bibinfo{pages}{47} (\bibinfo{year}{2017}), \eprint{1703.08521}.

\bibitem[{\citenamefont{Seikel et~al.}(2012)\citenamefont{Seikel, Clarkson, and
  Smith}}]{seikel2012reconstruction}
\bibinfo{author}{\bibfnamefont{M.}~\bibnamefont{Seikel}},
  \bibinfo{author}{\bibfnamefont{C.}~\bibnamefont{Clarkson}}, \bibnamefont{and}
  \bibinfo{author}{\bibfnamefont{M.}~\bibnamefont{Smith}},
  \bibinfo{journal}{Journal of Cosmology and Astroparticle Physics}
  \textbf{\bibinfo{volume}{2012}}, \bibinfo{pages}{036} (\bibinfo{year}{2012}).

\bibitem[{\citenamefont{Zhang and Xia}(2016)}]{zhang2016test}
\bibinfo{author}{\bibfnamefont{M.-J.} \bibnamefont{Zhang}} \bibnamefont{and}
  \bibinfo{author}{\bibfnamefont{J.-Q.} \bibnamefont{Xia}},
  \bibinfo{journal}{Journal of Cosmology and Astroparticle Physics}
  \textbf{\bibinfo{volume}{12}}, \bibinfo{pages}{005} (\bibinfo{year}{2016}).

\bibitem[{\citenamefont{Zhang and Xia}(2017)}]{zhang2017physical}
\bibinfo{author}{\bibfnamefont{M.-J.} \bibnamefont{Zhang}} \bibnamefont{and}
  \bibinfo{author}{\bibfnamefont{J.-Q.} \bibnamefont{Xia}},
  \bibinfo{journal}{arXiv:1701.04973}  (\bibinfo{year}{2017}).

\bibitem[{\citenamefont{Wasserman}(2006)}]{wasserman2006all}
\bibinfo{author}{\bibfnamefont{L.}~\bibnamefont{Wasserman}},
  \emph{\bibinfo{title}{All of Nonparametric Statistics}}
  (\bibinfo{publisher}{Springer Science \& Business Media},
  \bibinfo{year}{2006}).

\bibitem[{\citenamefont{Huterer and
  Starkman}(2003)}]{huterer2002parameterization}
\bibinfo{author}{\bibfnamefont{D.}~\bibnamefont{Huterer}} \bibnamefont{and}
  \bibinfo{author}{\bibfnamefont{G.}~\bibnamefont{Starkman}},
  \bibinfo{journal}{Phys. Rev. Lett.} \textbf{\bibinfo{volume}{90}},
  \bibinfo{pages}{031301} (\bibinfo{year}{2003}).

\bibitem[{\citenamefont{Zheng et~al.}(2014)\citenamefont{Zheng, Li, Li, Xia,
  Li, and Lu}}]{zheng2014constraints}
\bibinfo{author}{\bibfnamefont{W.}~\bibnamefont{Zheng}},
  \bibinfo{author}{\bibfnamefont{S.-Y.} \bibnamefont{Li}},
  \bibinfo{author}{\bibfnamefont{H.}~\bibnamefont{Li}},
  \bibinfo{author}{\bibfnamefont{J.-Q.} \bibnamefont{Xia}},
  \bibinfo{author}{\bibfnamefont{M.}~\bibnamefont{Li}}, \bibnamefont{and}
  \bibinfo{author}{\bibfnamefont{T.}~\bibnamefont{Lu}},
  \bibinfo{journal}{Journal of Cosmology and Astroparticle Physics}
  \textbf{\bibinfo{volume}{2014}}, \bibinfo{pages}{030} (\bibinfo{year}{2014}).

\bibitem[{\citenamefont{Neben and Turner}(2013)}]{neben2013beyond}
\bibinfo{author}{\bibfnamefont{A.~R.} \bibnamefont{Neben}} \bibnamefont{and}
  \bibinfo{author}{\bibfnamefont{M.~S.} \bibnamefont{Turner}},
  \bibinfo{journal}{Astrophys. J.} \textbf{\bibinfo{volume}{769}},
  \bibinfo{pages}{133} (\bibinfo{year}{2013}).

\bibitem[{\citenamefont{Demianski et~al.}(2012)\citenamefont{Demianski,
  Piedipalumbo, Rubano, and Scudellaro}}]{demianski2012high}
\bibinfo{author}{\bibfnamefont{M.}~\bibnamefont{Demianski}},
  \bibinfo{author}{\bibfnamefont{E.}~\bibnamefont{Piedipalumbo}},
  \bibinfo{author}{\bibfnamefont{C.}~\bibnamefont{Rubano}}, \bibnamefont{and}
  \bibinfo{author}{\bibfnamefont{P.}~\bibnamefont{Scudellaro}},
  \bibinfo{journal}{Monthly Notices of the Royal Astronomical Society}
  \textbf{\bibinfo{volume}{426}}, \bibinfo{pages}{1396} (\bibinfo{year}{2012}).

\bibitem[{\citenamefont{Maor et~al.}(2001)\citenamefont{Maor, Brustein, and
  Steinhardt}}]{maor2001limitations}
\bibinfo{author}{\bibfnamefont{I.}~\bibnamefont{Maor}},
  \bibinfo{author}{\bibfnamefont{R.}~\bibnamefont{Brustein}}, \bibnamefont{and}
  \bibinfo{author}{\bibfnamefont{P.~J.} \bibnamefont{Steinhardt}},
  \bibinfo{journal}{Physical Review Letters} \textbf{\bibinfo{volume}{86}},
  \bibinfo{pages}{6} (\bibinfo{year}{2001}).

\bibitem[{\citenamefont{Darling}(2012)}]{darling2012toward}
\bibinfo{author}{\bibfnamefont{J.}~\bibnamefont{Darling}},
  \bibinfo{journal}{Astrophys. J. Letters} \textbf{\bibinfo{volume}{761}},
  \bibinfo{pages}{L26} (\bibinfo{year}{2012}).

\bibitem[{\citenamefont{Steinmetz et~al.}(2008)\citenamefont{Steinmetz, Wilken,
  Araujo-Hauck, Holzwarth, H{\"a}nsch, Pasquini, Manescau, D'Odorico, Murphy,
  Kentischer et~al.}}]{steinmetz2008laser}
\bibinfo{author}{\bibfnamefont{T.}~\bibnamefont{Steinmetz}},
  \bibinfo{author}{\bibfnamefont{T.}~\bibnamefont{Wilken}},
  \bibinfo{author}{\bibfnamefont{C.}~\bibnamefont{Araujo-Hauck}},
  \bibinfo{author}{\bibfnamefont{R.}~\bibnamefont{Holzwarth}},
  \bibinfo{author}{\bibfnamefont{T.~W.} \bibnamefont{H{\"a}nsch}},
  \bibinfo{author}{\bibfnamefont{L.}~\bibnamefont{Pasquini}},
  \bibinfo{author}{\bibfnamefont{A.}~\bibnamefont{Manescau}},
  \bibinfo{author}{\bibfnamefont{S.}~\bibnamefont{D'Odorico}},
  \bibinfo{author}{\bibfnamefont{M.~T.} \bibnamefont{Murphy}},
  \bibinfo{author}{\bibfnamefont{T.}~\bibnamefont{Kentischer}},
  \bibnamefont{et~al.}, \bibinfo{journal}{Science}
  \textbf{\bibinfo{volume}{321}}, \bibinfo{pages}{1335} (\bibinfo{year}{2008}).

\bibitem[{\citenamefont{Chevallier and
  Polarski}(2001)}]{chevallier2001accelerating}
\bibinfo{author}{\bibfnamefont{M.}~\bibnamefont{Chevallier}} \bibnamefont{and}
  \bibinfo{author}{\bibfnamefont{D.}~\bibnamefont{Polarski}},
  \bibinfo{journal}{International Journal of Modern Physics D}
  \textbf{\bibinfo{volume}{10}}, \bibinfo{pages}{213} (\bibinfo{year}{2001}).

\bibitem[{\citenamefont{Linder}(2003)}]{linder2003exploring}
\bibinfo{author}{\bibfnamefont{E.~V.} \bibnamefont{Linder}},
  \bibinfo{journal}{Physical Review Letters} \textbf{\bibinfo{volume}{90}},
  \bibinfo{pages}{091301} (\bibinfo{year}{2003}).

\bibitem[{\citenamefont{{Jassal} et~al.}(2005)\citenamefont{{Jassal}, {Bagla},
  and {Padmanabhan}}}]{2005MNRAS.356L..11J}
\bibinfo{author}{\bibfnamefont{H.~K.} \bibnamefont{{Jassal}}},
  \bibinfo{author}{\bibfnamefont{J.~S.} \bibnamefont{{Bagla}}},
  \bibnamefont{and}
  \bibinfo{author}{\bibfnamefont{T.}~\bibnamefont{{Padmanabhan}}},
  \bibinfo{journal}{Mon. Not. R. Astron. Soc.} \textbf{\bibinfo{volume}{356}},
  \bibinfo{pages}{L11} (\bibinfo{year}{2005}), \eprint{astro-ph/0404378}.

\bibitem[{\citenamefont{Nair et~al.}(2014)\citenamefont{Nair, Jhingan, and
  Jain}}]{nair2014exploring}
\bibinfo{author}{\bibfnamefont{R.}~\bibnamefont{Nair}},
  \bibinfo{author}{\bibfnamefont{S.}~\bibnamefont{Jhingan}}, \bibnamefont{and}
  \bibinfo{author}{\bibfnamefont{D.}~\bibnamefont{Jain}},
  \bibinfo{journal}{Journal of Cosmology and Astro-Particle Physics}
  \textbf{\bibinfo{volume}{1}}, \bibinfo{pages}{005} (\bibinfo{year}{2014}).

\bibitem[{\citenamefont{Crittenden et~al.}(2012)\citenamefont{Crittenden, Zhao,
  Pogosian, Samushia, and Zhang}}]{crittenden2012fables}
\bibinfo{author}{\bibfnamefont{R.~G.} \bibnamefont{Crittenden}},
  \bibinfo{author}{\bibfnamefont{G.-B.} \bibnamefont{Zhao}},
  \bibinfo{author}{\bibfnamefont{L.}~\bibnamefont{Pogosian}},
  \bibinfo{author}{\bibfnamefont{L.}~\bibnamefont{Samushia}}, \bibnamefont{and}
  \bibinfo{author}{\bibfnamefont{X.}~\bibnamefont{Zhang}},
  \bibinfo{journal}{Journal of Cosmology and Astroparticle Physics}
  \textbf{\bibinfo{volume}{2012}}, \bibinfo{pages}{048} (\bibinfo{year}{2012}).

\bibitem[{\citenamefont{Ma and Corasaniti}(2016)}]{ma2016statistical}
\bibinfo{author}{\bibfnamefont{C.}~\bibnamefont{Ma}} \bibnamefont{and}
  \bibinfo{author}{\bibfnamefont{P.-S.} \bibnamefont{Corasaniti}},
  \bibinfo{journal}{arXiv:1604.04631}  (\bibinfo{year}{2016}).

\end{thebibliography}
\end{document}